\documentclass[prd,aps,showpacs,nofootinbib,preprintnumbers,onecolumn]{revtex4-1}
\usepackage{amsfonts,amsmath,amssymb,bm,dsfont,eucal}
\usepackage[english]{babel}
\usepackage[latin1]{inputenc}
\usepackage[T1]{fontenc}
\usepackage{ae,aecompl}
\usepackage{graphicx,graphics,epsfig,epstopdf,subfigure,color,psfrag}% Include figure files
\usepackage{hyperref}
\newcommand{\be}{\begin{equation}}
\newcommand{\ee}{\end{equation}}
\newcommand{\bea}{\begin{eqnarray}}
\newcommand{\eea}{\end{eqnarray}}

\newcommand{\ket}[1]{\left| #1 \right\rangle}
\begin{document}

\preprint{BARI-TH/647-12}
\title{ $B\to K \eta^{(\prime)} \gamma$ decays in the standard model \\and in  scenarios with universal extra dimensions}
\author{P. Biancofiore$^{a,b}$, P. Colangelo$^b$ and  F. De Fazio$^b$}
\affiliation{
$^a$Dipartimento di Fisica,  Universit$\grave{a}$  di Bari, Italy\\
$^b$Istituto Nazionale di Fisica Nucleare, Sezione di Bari, Italy}

\begin{abstract}
We study the radiative $B \to K \eta^{(\prime)} \gamma$  decays,  which are important  to investigate CP violation, and are also relevant to assess the role of the exclusive modes induced by the $b \to s \gamma$ transition to saturate the inclusive $B \to X_s \gamma$ decay rate. Moreover,
these channels do not display the same hierarchy as  $B \to K \eta^{(\prime)}$ modes, for which  the decay into $\eta^\prime$  is enhanced with respect to one into $\eta$. The three-body radiative decays reverse the role:  we find that this experimentally observed behavior   (although affected by a large uncertainty in the case of the $\eta^\prime$)   is reproduced in the theoretical analysis. We compute a   $B^* \to K$ form factor, needed for this study,  using light cone QCD sum rules, and discuss a  relation expected to hold in  the large energy  limit for the light meson. Finally, we  examine $B\to K \eta \gamma$ in two extensions of the standard model with universal extra dimensions, to investigate the sensitivity of this rare mode to such a kind of new physics effects.
\end{abstract}

 \maketitle

\section{Introduction}
The decay processes  driven by the flavor changing neutral current (FCNC)  $b \to s $ transitions provide efficient tests of the standard model (SM) and can display deviations signaling  new physics (NP)
phenomena.   Among such processes,  the  $b \to s $   induced decays of $B$ mesons are the best studied and  experimentally investigated;   several of them  have been observed through dedicated experimental analyses which have produced measurements of   a variety of observables useful, on the basis of information on  both inclusive and exclusive channels,  to confirm  the SM and  constrain  NP  scenarios \cite{Buchalla:2008jp}.

The radiative $b \to s \gamma$ transition,  on which we focus here, is particularly relevant.  Branching fractions have been measured for the inclusive $B \to X_s \gamma$ mode, as well as for several exclusive channels,  namely $B \to K^*(892) \gamma$, $B \to K_1(1270) \gamma$, $B \to K^*_2(1430) \gamma$, $B \to K \eta \gamma$, $B \to K \eta^\prime \gamma$, $B \to K \phi \gamma$, $ B \to K^*(892) \pi \gamma$ and $B \to K \pi \pi \gamma$, both for neutral and charged $B$ mesons \cite{pdg}. The  observed exclusive modes do not saturate the inclusive rate, therefore the scrutiny of the exclusive transitions is mandatory  in view of understanding the hadronization process  for  this class of channels.
This is one of the  motivations of the  present analysis of  the three-body $B \to K \eta^{(\prime)} \gamma$ modes.  Moreover,
there are other features making the multibody decays  induced by $b \to s \gamma$ interesting  to be studied. First,  the time-dependent CP asymmetry in  the neutral modes $ B^0 \to K^0_{S,L} \eta^{(\prime)} \gamma$ is sensitive to NP,  which may also manifest itself in producing right-handed photons; indeed,   in the SM the  photons produced in the $b \to s \gamma$ transition are mainly left-handed,  the amplitude for
emitting right-handed photons being suppressed by the quark mass ratio $m_s/m_b$   \cite{hep-ph/0410036}.
Furthermore,   the branching fractions of $B \to K \eta \gamma$ and $B \to K \eta^\prime \gamma$ do not obey  the same hierarchy as in the two-body decays  $B \to K \eta$ and $B \to K \eta^\prime$,    the last process being enhanced with respect to the former one.  The enhancement of two-body hadronic transitions with $\eta^\prime$ in the final state is common to several $B$ and $D$ decays,  and is not yet fully understood. In the case of $D_s \to \eta^{(\prime)} \pi \, , \,\, \eta^{(\prime)} \rho$,  the gluon content of the $\eta^\prime$ has been  indicated as  playing an important  role \cite{Colangelo:2001cv}.
For $B \to K \eta$ and $B \to K \eta^\prime$,  a possible explanation of the hierarchy between the two decay rates has been found in the destructive interference  among the penguins contributions   \cite{Lipkin:1990us},
 and, modulo large uncertainties, this has been numerically reproduced in the framework of QCD factorization \cite{Beneke:2002jn}. On the contrary, the radiative modes
  $B \to K \eta \gamma$ and $B \to K \eta^\prime \gamma$ show the opposite trend, as one can infer from   the  results   provided by
 Belle \cite{Nishida:2004fk,arXiv:0810.0804}  and BaBar collaborations \cite{Aubert:2006vs,Aubert:2008js},  and  collected in Table \ref{table:br}:  such an outcome deserves investigations.
%%%%%%%%%%%%%%%%%%%%%%%%%%%%%%%%%%%%%%%%%%%%
\begin{table*}[h!]
\centering \caption{Experimental results for the $B\to K \eta^{(\prime)} \gamma$ branching fractions ($\times 10^6$) from Belle and BaBar. The upper limits are at $90\%$ CL.}\label{table:br}
   \begin{ruledtabular}
\begin{tabular}{c  c c  }
Mode &  Belle collaboration\,\,\,\,\,\,\,\,\, & BaBar collaboration \hspace*{0.6cm} \\ \hline
$B^+ \to K^+ \, \eta \, \gamma$ & $8.4 \pm 1.5 \pm^{1.2}_{0.9}$ \,\,\,\, \cite{Nishida:2004fk} & $7.7 \pm 1.0 \pm 0.4 $  \,\,\,\,\,\,\,  \cite{Aubert:2008js} \\ \hline
$B^0 \to K^0 \, \eta \, \gamma$ & $8.7 \pm^{3.1}_{2.7} \pm^{1.9}_{2.6}$  \,\,\,\,\,  \cite{Nishida:2004fk} & $7.1 \pm^{2.1}_{2.0} \pm 0.4 $  \,\,\,\,\,\,\,\,\,\,  \cite{Aubert:2008js} \\ \hline
$B^+ \to K^+ \, \eta^\prime \, \gamma$ & $3.6 \pm 1.2 \pm 0.4$  \, \cite{arXiv:0810.0804} & $1.9 \pm^{1.5}_{1.2} \pm 0.1 $  \,\,\,($< 4.2$) \cite{Aubert:2006vs}\\ \hline
$B^0 \to K^0 \, \eta^\prime \, \gamma$ &  \hspace*{0.5cm} $<6.4$ \hspace*{0.7cm} \cite{arXiv:0810.0804} & $1.1 \pm^{2.8}_{2.0} \pm 0.1 $   \,\,\,($< 6.6$)   \cite{Aubert:2006vs}\\
   \end{tabular}
\end{ruledtabular}
\end{table*}
From the experimental side, the BaBar collaboration has also measured  the mixing induced ($S$) and direct  ($C$) CP asymmetries in the  $B^0 \to K_S^0 \eta \gamma$ transition. At present they are both compatible with zero:
$\displaystyle S=-0.18 \pm^{0.49}_{0.46} \pm 0.12$ and $\displaystyle C=-0.32 \pm^{0.40}_{0.39} \pm 0.07$   \cite{Aubert:2008js}.

The processes $B \to K \eta^{(\prime)} \gamma$ have  been studied in Ref.\cite{Fajfer:2008zy}  considering exclusively the regions of the phase space where one of the two pseudoscalar mesons in the final state is soft,  while the photon is hard. Describing  the amplitudes   as taking contributions only  from
 virtual intermediate $B^*$  and  $B_s^*$,  the Heavy Quark Effective Theory together  with the light meson chiral perturbation theory ($\chi$HQET) has been employed to describe the decays
 in   corners of the Dalitz plot; moreover, the $\eta-\eta^\prime$ mixing has been described in  the octet-singlet  mixing scheme. As a  result,  a fraction of about $10\%$ of the measured  $B \to K \eta \gamma$ branching ratio has been obtained.

In the present study  we  improve the analysis  in many respects.
We take into account several  possible underlying transitions, depicted in Fig. \ref{diagrams}, observing that, in addition to $b \to s \gamma$,  the transition $b \to s \bar q q$ can contribute to the processes.
In particular,  the explicit calculation shows that the diagram (1) with intermediate virtual $K^*$ is important,
together with diagrams (3) and (4),  while the one with intermediate $K^*_2(1430)$  and the other diagrams in Fig. \ref{diagrams} are smaller.
Furthermore, we do not confine ourselves to portions of the phase space, but  we extend the study to  the full three-body Dalitz plot. This requires  new information, namely the
$B^* \to K$ and $B_s^* \to \eta$ form factors,  which we compute by light cone QCD sum rules for a physical value of the beauty quark mass and for a wide range of  four-momentum transferred.
Finally,  we  consider the $\eta-\eta^\prime$ system in the  flavor basis, in which the mixing is described by a  single mixing angle  experimentally
 determined with  high accuracy by the KLOE collaboration from radiative $\phi \to \eta^{(\prime)} \gamma$ decay data.   In this way,  results for the SM can be obtained, and  the effects of  NP extensions, such as   scenarios with  universal extra dimensions (UED), can be examined.

All these topics are described  in the forthcoming  sections. In particular,
in Section \ref{calc}  we set up the stage of our calculation, considering the  diagrams  taken  into account in the theoretical description of $B \to K \eta^{(\prime)} \gamma$.
We provide the expressions of the various amplitudes and identify  the quantities  necessary for  their evaluation. Section \ref{LCSR} is devoted to  the light cone QCD sum rule determination of the form factor $T_1^{B^* \to K}$, which enters in the analysis of  $B \to K \eta^{(\prime)} \gamma$ transitions;  we collect  in the Appendix the definitions of the needed kaon light-cone distribution amplitudes (LCDA).
 The  computation of the decay rates is carried out in Sec.\ref{numerics} in the SM; the sensitivity to NP  effects  of one and two universal extra dimension scenarios is also investigated.
Section \ref{conclusion} contains our conclusions.

\section{The decays $B \to K \eta^{(\prime)} \gamma$}\label{calc}

\begin{figure}[b]
\includegraphics[scale=0.6]{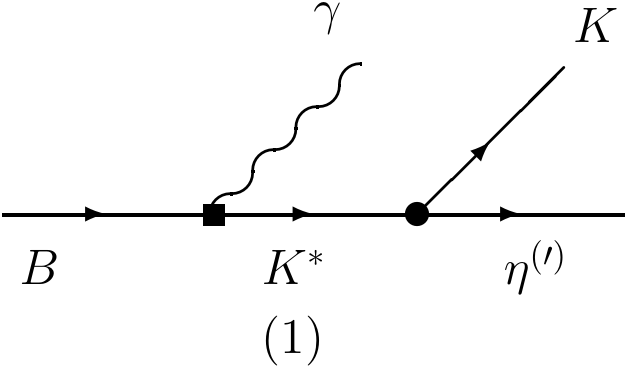}%
\hspace*{1.5cm}
\includegraphics[scale=0.6]{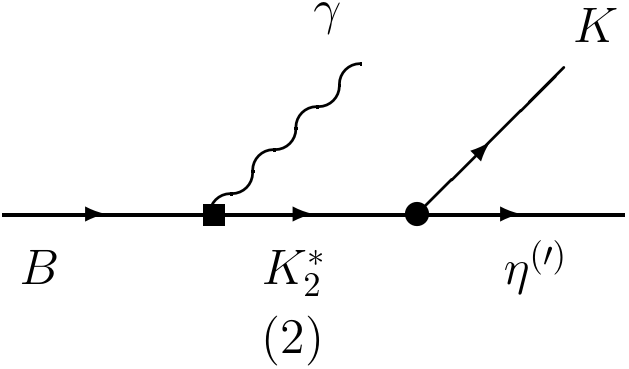}%
\hspace*{1.5cm}
\includegraphics[scale=0.6]{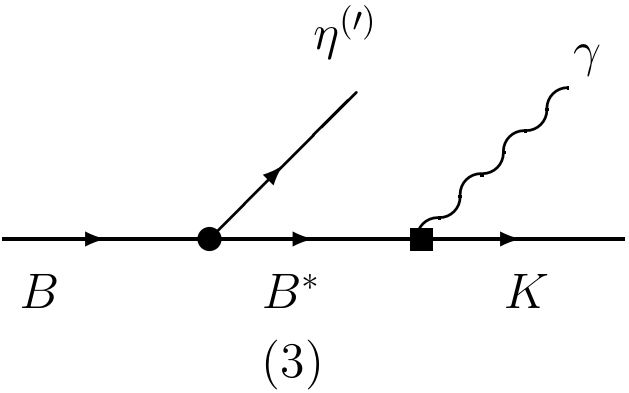}%
\\
\vspace*{0.5cm}
\includegraphics[scale=0.6]{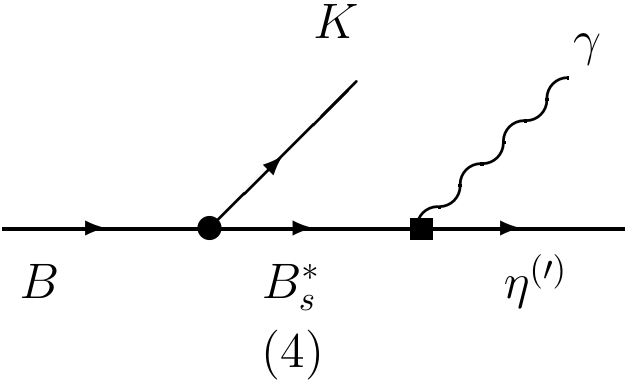}%
\hspace*{1.5cm}
\includegraphics[scale=0.6]{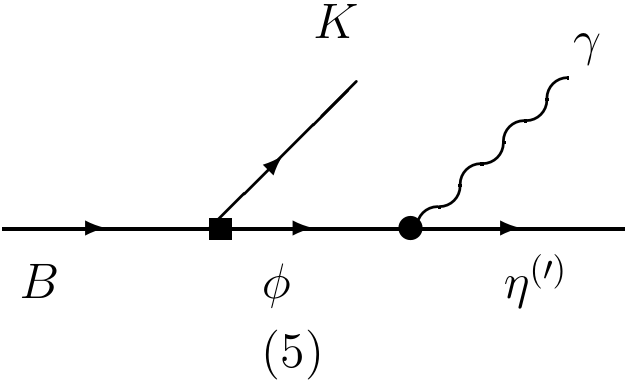}%
\hspace*{1.5cm}
\includegraphics[scale=0.6]{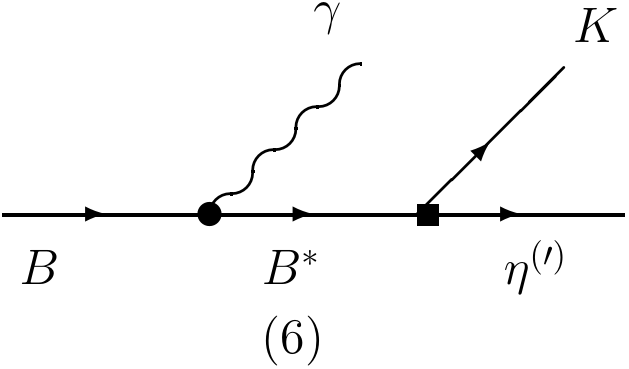}%
 \caption{Some diagrams contributing to the decays ${\overline B^0} \to {\overline K^0} \eta^{(\prime)} \gamma$. The dots indicate em and strong couplings; the squares indicate weak vertices. }\label{diagrams}
\end{figure}

We consider the transitions ${\overline B^0} (p) \to {\overline K^0}(p_K) \, \eta^{(\prime)}(p_{\eta^{(\prime)}}) \, \gamma(q,\epsilon)$, where $p$, $p_K$ and $p_{\eta^{(\prime)}}$ are the four momenta of  $B$, $K$ and $\eta^{(\prime)}$, respectively,
while $q$ and $\epsilon$ are the photon four momentum and  polarization vector. Although we refer to the decays of the neutral  ${\overline B^0}$,  in the following we  omit the charge adopting a simpler notation;
at the end of our study we shall comment  on the charged $B$ meson decays.
The three-body transitions can be described as proceeding through intermediate states: The ones that we take into account are displayed in Fig. \ref{diagrams}.
The first two diagrams (1) and (2) take contribution from intermediate $K^*(892)$ and $K_2^*(1430)$, respectively,  which have
width: $\Gamma(K^{*0}(892))=48.7 \pm 0.8$ MeV and $\Gamma(K^*_2(1430))=98.5 \pm 2.9$ MeV \cite{pdg}.  Higher kaon excitations are expected to give a smaller contribution, due to their larger widths
and to  the suppression provided by their propagators in the corresponding diagrams.
%
%%%%%%%%%%%%%%%%%%%%%%%%%%%%%%%%%%%%%%%%%%%%%%%%%%%%%
%%%%%%%%%%%%%%%%%%%%%%%%%%%%%%%%%%%%%%%%%%%%%%%%%%%%%
%
The  two diagrams (3) and (4)  have $B^*$ or $B_s^*$ as intermediate states, which are very narrow so that we neglect their widths.  The diagram (5) takes contribution from the intermediate
$\phi(1020)$ decaying to $\eta^{(\prime)} \gamma$ and having  $\Gamma(\phi(1020))=4.26 \pm 0.04$ MeV;
$B^*$ is the intermediate state also in the last diagram, which involves the radiative  $B^*B\gamma$ vertex.

To calculate the  amplitudes corresponding to the diagrams in Fig. \ref{diagrams}  we  need the effective weak Hamiltonian describing the
 $b \to s \gamma$ and $b \to s \, {\rm gluon}$ transition.  In the SM this reads \cite{Buchalla:1995vs}:
\be
H_{eff}=-{G_F \over \sqrt{2}} V_{tb}V_{ts}^* \left( \sum_{i=1}^6 C_i \, O_i+C_{7 \gamma} \, O_{7 \gamma}+C_{8G} \,  O_{8G} \right) \,. \label{heff}
\ee
$G_F$ is the Fermi constant and $V_{ij}$ are elements of the Cabibbo-Kobayashi-Maskawa mixing matrix; we
neglect terms proportional to $V_{ub} V_{us}^*$ since the ratio
$\displaystyle \left |{V_{ub}V_{us}^* \over V_{tb}V_{ts}^*}\right|$ is of ${O}(10^{-2})$.
$C_i$ are Wilson coefficients, while $O_i$ are local operators written in
terms of quark and gluon fields:
\begin{eqnarray}
O_1&=&({\bar s}_{L \alpha} \gamma^\mu b_{L \alpha})
      ({\bar c}_{L \beta} \gamma_\mu c_{L \beta}) \nonumber \\
O_2&=&({\bar s}_{L \alpha} \gamma^\mu b_{L \beta})
      ({\bar c}_{L \beta} \gamma_\mu c_{L \alpha}) \nonumber \\
O_3&=&({\bar s}_{L \alpha} \gamma^\mu b_{L \alpha})
      [({\bar u}_{L \beta} \gamma_\mu u_{L \beta})+...+
      ({\bar b}_{L \beta} \gamma_\mu b_{L \beta})] \nonumber \\
O_4&=&({\bar s}_{L \alpha} \gamma^\mu b_{L \beta})
      [({\bar u}_{L \beta} \gamma_\mu u_{L \alpha})+...+
      ({\bar b}_{L \beta} \gamma_\mu b_{L \alpha})] \nonumber \\
O_5&=&({\bar s}_{L \alpha} \gamma^\mu b_{L \alpha})
      [({\bar u}_{R \beta} \gamma_\mu u_{R \beta})+...+
      ({\bar b}_{R \beta} \gamma_\mu b_{R \beta})]  \\
O_6&=&({\bar s}_{L \alpha} \gamma^\mu b_{L \beta})
      [({\bar u}_{R \beta} \gamma_\mu u_{R \alpha})+...+
      ({\bar b}_{R \beta} \gamma_\mu b_{R \alpha})] \nonumber \\
O_{7 \gamma}&=&{e \over 16 \pi^2} \left[ m_b ({\bar s}_{L \alpha}
\sigma^{\mu \nu}
     b_{R \alpha}) +m_s ({\bar s}_{R \alpha}
\sigma^{\mu \nu}
     b_{L \alpha}) \right] F_{\mu \nu} \nonumber \\
O_{8G}&=&{g_s \over 16 \pi^2} m_b \Big[{\bar s}_{L \alpha}
\sigma^{\mu \nu}
      \Big({\lambda^a \over 2}\Big)_{\alpha \beta} b_{R \beta}\Big] \;
      G^a_{\mu \nu} \,\,. \nonumber
\label{eff}
\end{eqnarray}
\noindent
$\alpha$, $\beta$ are color indices, $\displaystyle
b_{R,L}={1 \pm \gamma_5 \over 2}b$, and $\displaystyle \sigma^{\mu
\nu}={i \over 2}[\gamma^\mu,\gamma^\nu]$; $e$ and $g_s$ are the
electromagnetic and the strong coupling constant, respectively, $m_b$ and $m_s$ are the beauty and the strange quark masses, while
$F_{\mu \nu}$  in $O_{7 \gamma}$ and $G^a_{\mu \nu}$ in  $O_{8 \gamma}$ denote the
electromagnetic and the gluonic field strength tensors. $\lambda^a$ are the Gell-Mann matrices.

The Wilson coefficients appearing in (\ref{heff}) have been computed at next-to-next-to-leading order in the standard model  \cite{nnlo}.
The most relevant contribution to $b \to s \gamma$ comes from the operator $ O_{7 \gamma}$, which is a magnetic penguin specific of such a transition and originates from the mass insertion on the external $b$-quark line in the QED penguin. The term proportional to $m_s$ contributes much less than the one proportional to $m_b$, and this is the reason for which the emission of left-handed photons dominates over that of right-handed ones in the SM. Since   the coefficient $C_{7 \gamma}$ depends on the  regularization scheme,  it is convenient to consider at leading order a combination that is  regularization scheme independent \cite{Buras:1993xp}:
\begin{equation}
 C_{7\gamma}^{(0)eff}(\mu_b)=\eta^{16 \over 23}
C_{7\gamma}^{(0)}(\mu_W)+{8 \over 3} \left( \eta^{14 \over 23} -\eta^{16
\over 23} \right)C_{8G}^{(0)}(\mu_W)+C_2^{(0)}(\mu_W) \sum_{i=1}^8
h_i \eta^{a_i} \,\,\, , \ee
where $\displaystyle \eta={\alpha_s(\mu_W) \over
\alpha_s(\mu_b)}$ and $C_2^{(0)}(\mu_W)=1$
(the superscript $(0)$ stays for leading log approximation); furthermore,
\bea a_1={14 \over 23} \hskip 0.4cm a_2={16 \over 23}
\hskip 0.4cm && a_3={6 \over 23} \hskip 0.4cm a_4=-{12 \over 23}
\nonumber \\ a_5= 0.4086 \hskip 0.4cm a_6=-0.4230 \hskip 0.4cm &&
a_7=-0.8994 \hskip 0.4cm
a_8=0.1456 \nonumber \\
h_1=2.2996 \hskip 0.4cm h_2=-1.0880  \hskip 0.4cm &&
 h_3=-{3 \over 7}  \hskip 0.4cm  h_4=-{1 \over 14} \label{numbers}
 \\
 h_5=-0.6494 \hskip 0.4cm h_6=-0.0380 \hskip 0.4cm && h_7= -0.0185 \hskip 0.4cm
 h_8=-0.0057 \,\,\, . \nonumber \eea
The effective weak vertex $O_{7 \gamma}$ contributes  to the diagrams (1-4) in Fig. \ref{diagrams}  through  hadronic matrix elements that we define below.   However, before doing that,  we  turn to the
$\eta-\eta^\prime$ system.

The  $\eta-\eta^\prime$ mixing  is usually described in
two  different schemes,  adopting  either the singlet-octet  or  the quark flavor (QF) basis, and in each scheme  two mixing angles are involved \cite{Feldmann:1999uf}.
Here we adopt the quark flavor basis defining
 \bea \ket{\eta_q}&=&{1 \over \sqrt{2} } \left(
\ket{{\bar
u} u} +\ket{{\bar d} d}\right) \nonumber \\
\ket{\eta_s}&=& \ket{{\bar s} s} \,\, , \label{etaqs} \eea
so that the $\eta$-$\eta^\prime$ system can be described in terms of the  mixing angles  $\varphi_q$ and $\varphi_s$:
\bea
\ket{\eta}&=&  \cos \, \varphi_q \ket{\eta_q}- {\sin} \, \varphi_s  \ket{\eta_s} \nonumber \\
\ket{\eta^\prime}&=& { \sin} \, \varphi_q  \ket{\eta_q}+{\cos} \, \varphi_s \ket{\eta_s}  \,\,\,\, . \label{mixing}
\eea
The difference between  $\varphi_q$ and $\varphi_s$ is due to OZI-violating effects and is experimentally found to be small
($\varphi_q-\varphi_s  <  5^\circ$), so that it has been proposed that the approximation of describing the $\eta-\eta^\prime$ mixing in the QF basis and a single mixing angle
is  suitable \cite{Feldmann:1999uf}.
The simplification  $\varphi_q\simeq \varphi_s \simeq \varphi$ is  supported by a QCD sum rule analysis
of the  $\phi \to \eta\gamma$ and  $\phi \to \eta^\prime\gamma$ decays \cite{DeFazio:2000my}.
A precise determination of the $\eta-\eta^\prime$ mixing angle has been obtained by the KLOE Collaboration measuring
 the ratio $\displaystyle{ \Gamma(\phi \to \eta^\prime  \gamma) \over \Gamma(\phi \to \eta \gamma)}$ in the flavor basis  with  a single mixing angle, with the result:
$\varphi=\big( 41.5 \pm 0.3_{stat} \pm 0.7_{syst} \pm0.6_{th} \big )^\circ$  \cite{kloe}.
This analysis has been  improved performing a global fit of the transitions $V \to P \gamma$ and $P \to V \gamma$ ($V=\phi, \, \omega, \, \rho$ and $P=\pi^0,\,\eta, \, \eta^\prime$),  allowing a gluonium content in the $\eta^\prime$ and  including the measurement of the ratio $\displaystyle {\Gamma (\eta^\prime \to \gamma \gamma) \over \Gamma(\pi^0 \to \gamma \gamma)}$   \cite{Ambrosino:2009sc}. The outcome is that, even though  the gluonium content of the $\eta^\prime$ is significant, the result for the $\eta-\eta^\prime$ mixing angle is only negligibly affected.  Therefore, we set $\varphi$  to the value quoted above.

Let us now consider in turn the various diagrams in Fig. \ref{diagrams}.
\begin{itemize}
\item Diagrams 1 and 2
\end{itemize}
The corresponding amplitudes read:
\bea
A_1 &=& A(B \to K^* \gamma) {i \over s-m_{K^*}^2+i\,m_{K^*} \, \Gamma_{K^*}}A(K^* \to K \eta^{(\prime)})
\label{a1} \\
A_2 &=& A(B \to K^*_2 \gamma) {i \over s-m_{K^*_2}^2+i\,m_{K^*_2} \, \Gamma_{K^*_2}}A(K^*_2 \to K \eta^{(\prime)})
\label{a2}\,\,,
\eea
with
\bea
A(B \to K^*_{(2)} \gamma)=C\epsilon^{*\mu} \left[(m_b+m_s)<K^*_{(2)}(p_K,{\tilde \epsilon})|{\bar s} \sigma_{\mu \nu} q^\nu
  b |B(p)>+(m_b-m_s)<K^*_{(2)}(p_K,{\tilde \epsilon})|{\bar s} \sigma_{\mu \nu} q^\nu
 \gamma_5  b |B(p)> \right]  \,\,\, , \nonumber
\eea
to be computed for an on-shell  ($q^2=0$) photon, defining  $s=(p-q)^2=M^2_{K \eta^{(\prime)}}$.
 The factor $C$ is  $C=4\displaystyle{G_F \over \sqrt{2}}V_{tb}V_{ts}^* C_7^{(eff)}\displaystyle{e \over 16 \pi^2}$.  The hadronic matrix elements of weak Hamiltonian operators are parametrized  in terms of form factors:
\begin{eqnarray}
<K^*_{(2)}(p_K,{\tilde \epsilon})|{\bar s} \sigma_{\mu \nu} q^\nu
  b |B(p)>&=& i \epsilon_{\mu \nu \alpha
\beta} {\tilde \epsilon}^{* \nu} p^\alpha p^{\beta}_K
\; 2 \; T_1^{B \to K^*_{(2)}}(q^2)   \label{t1BK*}
\\
<K^*_{(2)}(p_K,{\tilde \epsilon})|{\bar s} \sigma_{\mu \nu} q^\nu
 \gamma_5  b |B(p)>&=&
  \Big[ {\tilde \epsilon}^*_\mu (M_B^2 - s)  -
({\tilde \epsilon}^* \cdot q) (p+p_K)_\mu \Big] \; T_2^{B \to K^*_{(2)}}(q^2) \nonumber \\
&+& ({\tilde \epsilon}^* \cdot q) \left [ q_\mu - {q^2 \over M_B^2 -
s} (p + p_K)_\mu \right ] \; T_3^{B \to K^*_{(2)}}(q^2)  \; \,\,\, ,   \label{t23BK*}
\end{eqnarray}
with ${\tilde \epsilon}$ denoting the polarization vector of the $K^*_{(2)}$ mesons; in the case of $K^*_2(1430)$, which is a spin 2 particle, the polarization is described by a two indices symmetric and traceless tensor, therefore in (\ref{t1BK*}-\ref{t23BK*})  it is understood that  ${\tilde \epsilon}^\alpha={\tilde \epsilon}^{\alpha \beta}\displaystyle{p_\beta \over M_B}$. The condition holds: $T_1^{B \to K^*_{(2)}}(0) =T_2^{B \to K^*_{(2)}}(0) $.
The variable $s$  in the definition of the hadronic matrix elements takes into account  that the $K^*_{(2)}$ mesons are off-shell, and is needed to ensure gauge invariant amplitudes.

In the same diagrams strong vertices also appear, which are  defined as follows:
\bea
A(K^* \to K \eta^{(\prime)})&=&g_{K^*K\eta^{(\prime)}}\, {\tilde \epsilon} \cdot p_{\eta^{(\prime)}} \label{g1} \\
A(K^*_2 \to K \eta^{(\prime)})&=&g_{K^*_2K\eta^{(\prime)}}\,{\tilde \epsilon}^{\alpha \beta}  p_{\eta^{(\prime)} \alpha} \, p_{\eta^{(\prime)} \beta} \label{g2} \,\,.
\eea
Within the flavor scheme for the $\eta-\eta^\prime$ mixing the relations  $g_{K^*K\eta}=(\cos \varphi +\sqrt{2} \sin \varphi) g_{K^{*+}K^+\pi^0}$ and $g_{K^*K\eta^\prime}=(\sin \varphi -\sqrt{2} \cos \varphi) g_{K^{*+}K^+\pi^0}$ can be worked out. Assuming  the width of $K^{*+}$  saturated by the two modes $K^{*+} \to K^+ \pi^0,\,K^0\pi^+$, and using the relation $g_{K^{*+}K^0\pi^+}=\sqrt{2}\,g_{K^{*+}K^+\pi^0}$, from  $\Gamma(K^{*+})=50.8\pm0.9$ MeV   we obtain $g_{K^{*+}K^+\pi^0}=6.5 \pm 0.06$.

The  strong coupling  $g_{K^*_2K\eta}$  can be estimated, although with a large uncertainty, using the measurement  ${\cal B}(K^*_2 \to K \eta)=(1.5 \pm^{3.4}_{1.0}) \times 10^{-3}$ \cite{pdg} together with  $\Gamma(K^*_2)$,   obtaining: $g_{K^*_2K\eta}=1.43 \pm 1.60$ GeV$^{-1}$. On the other hand, no information is available for  $g_{K^*_2K \eta^\prime}$;  however, since, as we shall see, the contribution of this diagram is small in the case
 of $\eta$, it is reasonable to neglect it also in the case of the $\eta^\prime$ in the final state.

\newpage
\begin{itemize}
\item Diagrams 3 and 4
\end{itemize}
The two amplitudes read:
\bea
A_3 &=& A(B \to B^* \eta^{(\prime)}) {i \over t-m_{B^*}^2} A(B^* \to K \gamma)
\label{a3} \\
A_4 &=& A(B \to B^*_s K) {i \over u-m_{B^*_s}^2}A(B^*_s \to \eta^{(\prime)} \gamma)
\label{a4} \,\,,
\eea
with
\bea
A(B^* \to K \gamma)&=&C\epsilon^{*\nu} \left[(m_b+m_s)<K(p_K)|{\bar s} \sigma_{\mu \nu} q^\nu
  b |B^*(p^\prime,{\tilde \epsilon})>+(m_b-m_s)<K(p_K)|{\bar s} \sigma_{\mu \nu} q^\nu
\gamma_5  b |B^*(p^\prime,{\tilde \epsilon})> \right] \,\,\, ,  \nonumber \\
A(B^*_s \to \eta^{(\prime)} \gamma)&=&C\epsilon^{*\nu} \left[(m_b+m_s)<\eta^{(\prime)}(p_{\eta^{(\prime)}})|{\bar s} \sigma_{\mu \nu} q^\nu
  b |B^*_s(p^\prime,{\tilde \epsilon})>+(m_b-m_s)<\eta^{(\prime)}(p_{\eta^{(\prime)}})|{\bar s} \sigma_{\mu \nu} q^\nu
\gamma_5  b |B^*_s(p^\prime,{\tilde \epsilon})> \right] ,\nonumber
\eea
and
\bea
<K(p_K)|{\bar s} \sigma_{\mu \nu} q^\mu
  b |B^*(p^\prime,{\tilde \epsilon})>&=& i \epsilon_{\nu \tau \alpha
\beta} {\tilde \epsilon}^{ \tau} p^{\prime \alpha} p^{\beta}_K
\; 2 \; T_1^{B^* \to K}(q^2)  \label{t1B*K}
\\
<K(p_K)|{\bar s} \sigma_{\mu \nu} q^\mu
\gamma_5  b |B^*(p^\prime,{\tilde \epsilon})>&=&
 \Big[ {\tilde \epsilon}_\nu (t-m_K^2)  -
({\tilde \epsilon} \cdot q) (p^\prime+p_K)_\nu \Big] \; T_2^{B^* \to K}(q^2) \nonumber \\
&+& ({\tilde \epsilon} \cdot q) \left [ q_\nu - {q^2 \over t-m_K^2} (p^\prime + p_K)_\nu \right ] \; T_3^{B^* \to K}(q^2)  \; ,  \label{t23B*K}
\end{eqnarray}
\begin{eqnarray}
<\eta^{(\prime)}(p_{\eta^{(\prime)}})|{\bar s} \sigma_{\mu \nu} q^\mu
  b |B^*_s(p^\prime,{\tilde \epsilon})>&=& i \epsilon_{\nu \tau \alpha
\beta} {\tilde \epsilon}^{ \tau} p^{\prime \alpha} p^{\beta}_{\eta^{(\prime)}}
\; 2 \; T_1^{B^*_s \to \eta^{(\prime)}}(q^2)   \label{t1B*s-etaV}
\\
<\eta^{(\prime)}(p_{\eta^{(\prime)}})|{\bar s} \sigma_{\mu \nu} q^\mu
\gamma_5  b |B^*_s(p^\prime,{\tilde \epsilon})>&=&   \Big[ {\tilde \epsilon}_\nu (u-m_{\eta^{(\prime)}}^2)  -
({\tilde \epsilon} \cdot q) (p^\prime+p_{\eta^{(\prime)}})_\nu \Big] \; T_2^{B^*_s \to \eta^{(\prime)}}(q^2) \nonumber \\
&+& ({\tilde \epsilon} \cdot q) \left [ q_\nu - {q^2 \over u-m_{\eta^{(\prime)}}^2} (p^\prime + p_{\eta^{(\prime)}})_\nu \right ] \; T_3^{B^*_s \to \eta^{(\prime)}}(q^2)  \; .   \label{t1B*s-etaA}
\end{eqnarray}
($p^\prime$, ${\tilde \epsilon}$) denote the four momentum and the polarization vector of the $B^*_{(s)}$; moreover  $T_1^{B^* \to K}(0)= T_2^{B^* \to K}(0)$ and the same for $\eta^{(\prime)}$. The variable  $t=(q+p_K)^2$ takes into account the off-shellness of the $B^*$ in  diagram (3), while
 the variable $u=(p_{\eta^{(\prime)}}+q)^2$ accounts for the off-shellness of the $B_s^*$ in diagram (4); obviously,  $s+t+u=M_B^2+m_K^2+m_{\eta^{(\prime)}}^2$.

As for the strong vertices appearing in the two amplitudes, we define
\bea
A(B \to B^* \eta^{(\prime)})&=& g_{B^*B\eta^{(\prime)}}\,\,\, {\tilde \epsilon}^* \cdot p_{\eta^{(\prime)}} \label{g3} \\
A(B \to B^*_s K)&=& g_{B^*_sBK}\,\,\, {\tilde \epsilon}^* \cdot p_K \label{g4} \,\,.
\eea
The two couplings $g_{B^*B\eta^{(\prime)}}$ and $g_{B^*_s B K}$  can be obtained,  invoking $SU(3)_F$  symmetry, from the analogous quantity $g_{B^*B\pi}$: $g_{B^*B\eta}=\cos \varphi \, g_{B^*B\eta_q}=\cos \varphi \, \, g_{B^*B\pi} $, $g_{B^*B\eta\prime}=\sin \varphi \, g_{B^*B\eta_q}=\sin \varphi \, \, g_{B^*B\pi} $ and $g_{B^*_s B K}=g_{B^*B\pi}$.
As for $g_{B^*B\pi}$, it can be related to a low-energy parameter $g$ that describes the coupling of heavy mesons belonging to the doublet  of heavy-light quark states with spin-parity $J^P=(0^-,1^-)$ to light pseudoscalar states in the framework of the $\chi$HQET \cite{rev}:   $g_{B^*B\pi}={2M_B \over f_\pi}g$. There are several theoretical determinations of $g$ spanning the range $[0.2, \, 0.5]$ \cite{Colangelo:1995ph}. However, $g$ can be extracted from the measured decay width of $D^{*+} \to D^0 \pi^+$ \cite{Anastassov:2001cw},  obtaining $g=0.59 \pm 0.01 \pm 0.07$ \cite{Colangelo:2002dg}. We use this value in our analysis.
\begin{itemize}
\item Diagram 5
\end{itemize}
The contribution of the intermediate $\phi(1020)$ is represented by the amplitude
\be
A_5=A(B \to K \phi) {i \over u-m_\phi^2 +i\, m_\phi \, \Gamma_\phi}A(\phi \to \eta^{(\prime)} \gamma)
\label{diag5}\,\,. \ee
Adopting factorization, the first amplitude in (\ref{diag5}) can be written as
\be
A(B \to K \phi) ={G_F \over \sqrt{2}}V_{tb} V_{ts}^* a_w \langle K(p_K)| {\bar s} \gamma_\mu (1-\gamma_5) b |B(p) \rangle \, \langle \phi(p_\phi, {\tilde \epsilon}) |  {\bar s} \gamma_\mu  s |0 \rangle
\,\,, \label{bkphi}
\ee
where $a_w$ is an effective Wilson coefficient that we set to the value $a_w=0.064 \pm 0.009$ from the experimental branching fraction ${\cal B}({\overline B}^0 \to {\overline K}^0 \phi)=(8.6 \pm^{1.3}_{1.1})\times 10^{-6}$ \cite{pdg}.
Furthermore, we use the  parametrizations
\bea
\langle K(p_K)| {\bar s} \gamma_\mu (1-\gamma_5) b |B(p) \rangle &=& f_+^{B \to K}(q^2)(p+p_K)_\mu + f_-^{B \to K}(q^2)(p-p_K)_\mu \nonumber \\
\langle \phi(p_\phi {\tilde \epsilon}) |  {\bar s} \gamma_\mu  s |0 \rangle &=&
f_\phi m_\phi {\tilde \epsilon}^*_\mu \,\,. \label{bkphi-fact}
\eea
When these two  definitions are inserted in (\ref{bkphi})  only the form factor $f_+^{B \to K}$ contributes, and  we adopt for it the determination  in Ref.\cite{Ball:2004ye}. The value $f_\phi=(0.232 \pm 0.002)$ GeV
comes  from the experimental datum ${\cal B}(\phi \to e^+ e^-)=(2.954\pm 0.030) \times 10^{-4}$.

Following  Ref.\cite{DeFazio:2000my}, the amplitudes $A(\phi \to \eta^{(\prime)} \gamma)$ can be written as
\be
A(\phi \to \eta^{(\prime)} \gamma)=-{e \over 3}F^{\phi \to \eta^{(\prime)} \gamma} (q^2) \epsilon_{\nu \alpha \beta \delta} \epsilon^{* \nu} (p_\phi)^\alpha (p_\eta)^\beta {\tilde \epsilon}^\delta \,\,.
\ee
The form factors $F^{\phi \to \eta^{(\prime)} \gamma} (q^2)$ were determined using QCD sum rules, providing  their values at $q^2=0$ (multiplied by the strange quark charge in units of $e$): $|g^{\phi \eta \gamma}|=\displaystyle{{1 \over 3} F^{\phi \to \eta \gamma} (0)}=(0.66 \pm 0.06)$ GeV$^{-1}$ and $|g^{\phi \eta^\prime \gamma}|=\displaystyle{{1 \over 3}F^{\phi \to \eta^\prime \gamma} (0)}=(1.0 \pm 0.2)$ GeV$^{-1}$ \cite{DeFazio:2000my},  results used in our analysis.

\begin{itemize}
\item Diagram 6
\end{itemize}

It is  possible to show that the  diagram (6)  provides a tiny contribution with respect to the others. Let us discuss this in the case of the $\eta$.
The  amplitude can be written in terms of   $A(B^* \to B \gamma)$ and $A(B^* \to K \eta)$.
In order to understand how large this contribution  is, we can invoke naive factorization,  writing $A(B^* \to K \eta)={G_F \over \sqrt{2}} V_{ub}^* V_{us} a_2^{eff} \langle K| {\bar s} \gamma^\mu (1-\gamma_5)|  B^*\rangle \langle \eta | {\bar u} \gamma_\mu (1-\gamma_5) u |0 \rangle $, with $a_2^{eff}\simeq -0.286$  an effective Wilson coefficient for color suppressed decays.
The $\eta$-current-vacuum matrix element involves (in the flavor basis for the $\eta-\eta^\prime$ mixing) the constant $f_\eta^q=f_q \cos{\phi}$ with $f_q \simeq f_\pi$: $\langle \eta | {\bar u} \gamma_\mu (1-\gamma_5) u | 0 \rangle={i \over \sqrt{2}}  \, f_\eta^q \, (p_\eta)_\mu$. On the other hand,
the matrix element $\langle{K}| {\bar s} \gamma^\mu (1-\gamma_5) |B^* \rangle$  can be decomposed in terms of several form factors; however, when contracted with $ (p_\eta)_\mu$,  only one of such form factors contributes,  usually denoted as  $A_0(m_\eta^2)$, which  in the large energy limit  of the final light meson coincides with $T_1^{B^* \to K}$  computed in the next section.
The other ingredient  is the radiative amplitude $A(B^* \to B \gamma)$, which can be  written as  $A(B^*(p^\prime, {\tilde \epsilon}) \to B(p) \gamma(q,\epsilon))=e \left(\displaystyle{e_b \over \Lambda_b}+ \displaystyle{e_q \over \Lambda_q}\right) \, \epsilon_{\alpha \beta \tau \sigma}\, \epsilon^{* \alpha} \, {\tilde \epsilon}^\beta \, p^\tau \,p^{\prime \sigma}$, with  $e_b$ ($e_q$) the $b$ ($q=d$) quark charge in units of $e$.
A determination of the mass parameters $\Lambda_b$ and $\Lambda_q$ can be found in   \cite{Colangelo:1994jc}:   $\Lambda_b=4.93$ GeV (close to the $b$ quark mass) and $\Lambda_q=0.59$ GeV. As a result,  the contribution of the diagram (6)  to the branching fraction is ${\cal O}(10^{-13})$.   Therefore, in the following we  neglect this amplitude. \\

As it emerges from the above discussion, important quantities  are the form factors appearing in the  diagrams (1)-(4).  In the next section we compute
 $T_1^{B^* \to K}(q^2)$  by  light cone QCD sum rules \cite{Colangelo:2000dp}.  SU(3)$_F$ symmetry and the QF $\eta-\eta^\prime$ mixing scheme allow also to  fix: $T_1^{B^*_s \to \eta}(0)=-\sin \varphi \, T_1^{B^* \to K}(0)$ and $T_1^{B^*_s \to \eta^\prime}(0)=\cos \varphi \, T_1^{B^* \to K}(0)$.
As for $T_1^{B \to K^*}$,   several determinations can be found in the literature;  we use the  light cone QCD sum rule result $T_1^{B \to K^*}(0)=0.333 \pm 0.028$ \cite{Ball:2004rg},  to be consistent with the determination of $T_1^{B^* \to K}$. This value  is compatible with the one obtained by  three-point  QCD sum rules based on the short-distance expansion \cite{Colangelo:1995jv}.
Finally, for $T_1^{B \to K^*_2}$  we use $T_1^{B \to K^*_2}(0)=0.17 \pm 0.03 \pm 0.04$ \cite{Wang:2010ni}.

There is  a remark concerning the relative strong phases among the various amplitudes. While the sign between the amplitudes (3) and (4) can be fixed invoking $\chi$HQET and the flavor symmetry, the relative phase between, e.g., (1) and (3)
does not follow from theoretical arguments. Therefore, we consider it as a parameter to be determined empirically from the experimental data. The phases appearing in the other amplitudes do not play a role in the branching ratio due to the small size of such diagrams.

\section{Form factor $T_1^{B^* \to K}(q^2)$ by light cone QCD sum rules}\label{LCSR}

To compute the form factor $T_1^{B^* \to K}(q^2)$ by light cone QCD sum rules (LCSR) we  consider the two-point  correlation function with the external kaon state
\be
 \Pi_{\mu \nu}(p^\prime,q)= i \int d^4x \, e^{iq\cdot x} \langle K(p^\prime)|{\rm
 T}\left\{J_\mu(x),V_\nu(0)\right\}|0\rangle \,\,\,\,
 \label{corr}
\ee
where $J_\mu={\bar s} \sigma_{ \alpha\mu} q^\alpha b $ is the quark current appearing
in the  matrix element (\ref{t1B*K}).  $V_\nu={\bar b} \gamma_\nu q$ is the vector current with the quantum numbers of the $B^*$ meson ($q=u,\,d)$, and
its matrix element between the vacuum and the  $B^*$ state is parametrized in terms of the decay constant $f_{B^*}$,
\be
 \langle B^*(p^\prime+q,{\tilde \epsilon})| {\bar b} \gamma_\nu q|0\rangle =
 f_{B^*}m_{B^*}{\tilde \epsilon}^*_\nu \,\,.
\ee
The LCSR method consists in expressing the correlation
function  Eq.(\ref{corr}) both  in QCD and in terms of a hadronic representation.  $\Pi_{\mu \nu}$ can be decomposed in independent
 Lorentz structures, one of which can be used  to compute  $T_1^{B^* \to K}$:
\be
\Pi_{\mu \nu}(p^\prime,q)= i\,\, \epsilon_{\mu \alpha \tau \nu} q^\alpha p^{\prime \tau} \Pi((p^\prime+q)^2,q^2)+ {\rm other \,\, structures} \,\,.
\ee
In terms of hadronic states,  the correlation function in (\ref{corr}) can be written as
\be\label{had}
\Pi^{\rm HAD}_{\mu \nu}(p^\prime,q)=  \frac{\langle K(p^\prime)|J_\mu|B^*(p^\prime+q, {\tilde \epsilon})\rangle \langle B^*(p^\prime+q, {\tilde \epsilon})|V_\nu|0\rangle}
 {m_{B^*}^2-(p^\prime+q)^2}
 +\sum_h \frac{\langle K(p^\prime)|J_\mu|h(p^\prime+q)\rangle \langle h(p^\prime+q)|V_\nu|0\rangle}
 {m_h^2-(p^\prime+q)^2} \nonumber
\ee
and consists in  the contribution of the $B^*$ meson and of the
higher resonances and of the continuum of states $h$. The first term in (\ref{had})  contributes to the invariant function  $\Pi((p^\prime+q)^2,q^2)$, since
\be
\frac{\langle K(p^\prime)|J_\mu|B^*(p^\prime+q, {\tilde \epsilon})\rangle \langle B^*(p^\prime+q, {\tilde \epsilon})|V_\nu|0\rangle}
 {m_{B^*}^2-(p^\prime+q)^2}=\, i \, \epsilon_{\mu \alpha \tau \nu} q^\alpha p^{\prime \tau} \,\, {2 \,T_1^{B^* \to K}(q^2)\,f_{B^*} m_{B^*} \over m_{B^*}^2-(p^\prime+q)^2}
\,\,. \ee
 In a one-resonance+continuum  formulation, the hadronic representation of the function  $\Pi((p^\prime+q)^2,q^2)$
can be written as
\be
\Pi^{\rm HAD}((p^\prime+q)^2,q^2)= {2\, T_1^{B^* \to K}(q^2)\,f_{B^*} m_{B^*} \over m_{B^*}^2-(p^\prime+q)^2} + \int_{s_0}^\infty ds \frac{\rho^h(s,q^2)}
 {s-(p^\prime+q)^2}\, \,, \label{hadronic}
\ee
where  higher resonances and  the continuum of states are described in terms of the spectral function $\rho^h(s,q^2)$ which
contributes starting from a threshold $s_0$.

 The QCD expression of the correlation function is given by
\begin{eqnarray}
 \Pi^{\rm QCD}((p^\prime+q)^2,q^2)&=&   \frac{1}{\pi}\int_{m_b^2}^\infty ds \, \frac{{\rm Im}\Pi^{\rm QCD}(s,q^2)}
 {s-(p^\prime+q)^2} \,. \,\,\,\,\, \label{QCD-repr}
\end{eqnarray}
This expression comes from  an operator product expansion (OPE) of  the T-product in Eq.(\ref{corr}) on the light cone, which produces a series of operators, ordered by increasing twist, the matrix elements of which between the vacuum and the $K$ [required to evaluate   Eq.(\ref{corr})] are parametrized in terms of  $K$  light cone distribution amplitudes (LCDAs).
The equality  of the hadronic and QCD representations of the correlation function,  Eqs.(\ref{hadronic}) and (\ref{QCD-repr}), does not yet allow us  to derive  the $T_1^{B^* \to K}$ form factor,
since the hadronic spectral function $\rho^h$ is unknown. However, we can invoke global quark-hadron duality above the threshold $s_0$  \cite{shifman-duality}, which amounts to identify  integrals of the  spectral function $\rho^h$
with corresponding integrals of  $\rho^{\rm QCD}={1 \over \pi} {\rm Im} \Pi^{\rm QCD}$, and in particular
\begin{eqnarray}
\int_{s_0}^\infty  ds {\rho^h(s,q^2) \over s-(p^\prime+q)^2}&=&\frac{1}{\pi}\int_{s_0}^\infty ds \, \frac{{\rm Im}\Pi^{\rm QCD}(s,q^2)}{s-(p^\prime+q)^2}\,. \nonumber \\
\end{eqnarray}
Using global duality, together with  the equality $\Pi^{\rm HAD}_{\mu \nu}(p^\prime,q)=\Pi^{\rm QCD}_{\mu \nu}(p^\prime,q)$,
 from Eqs.(\ref{hadronic}) and (\ref{QCD-repr}) the equation follows:
\be
 {2 \,T_1^{B^* \to K}(q^2)\,f_{B^*} m_{B^*} \over m_{B^*}^2-(p^\prime+q)^2}= \frac{1}{\pi}\int_{m_b^2}^{s_0} ds \, \frac{{\rm Im}\Pi^{\rm QCD}(s,q^2)} {s-(p^\prime+q)^2} \,\,\ .
\label{res1}
\ee

The subtraction of the continuum and of the higher-twist contributions, leading to (\ref{res1}),  can be optimized, following the  QCD sum rule procedure, by a Borel transformation of the hadronic and of the QCD  expressions of the
correlation function,  hence of the two sides in Eq.(\ref{res1}). This transformation,  which applied to a function  ${\cal F}(Q^2)$  (with  $Q^2=-q^2$) is defined as
\be {\cal B} [{\cal
F}(Q^2)]=lim_{Q^2 \to \infty, \; n \to \infty, \; {Q^2 \over
n}=M^2}\; {1 \over (n-1)!} (-Q^2)^n \left({d \over dQ^2} \right)^n
{\cal F}(Q^2) \; , \label{tborel} \ee
where  $M^2$ is  the Borel parameter, produces the equality
 \be {\cal B} \left[ { 1 \over (s+Q^2)^n }
\right]={\exp(-s/M^2) \over (M^2)^n\ (n-1)!} \; . \label{bor} \ee
This operation improves the convergence of the OPE series  by
factorials of the power $n$,  and for suitably chosen values of $M^2$
enhances the contribution of the low lying states to the hadronic expression of the correlation function.
Applying the Borel transformation to both $\Pi^{\rm HAD}$ and $\Pi^{\rm QCD}$ we obtain:
\be
2\, T_1^{B^* \to K}(q^2)\,f_{B^*} m_{B^*}  \,\,
 {\rm
 exp}\left(-\frac{m_{B^*}^2}{M^2}\right)=\frac{1}{\pi}\int_{m_b^2}^{s_0}ds \,\,
 {\rm exp}\left(-\frac{s}{M^2}\right) \, \, {\rm Im}\Pi^{\rm QCD}(s,q^2)\,\,. \label{SR-generic}
\ee
The calculation of $\Pi^{\rm QCD}$,  based on the expansion of the
T-product in (\ref{corr}) near the light-cone,   involves
matrix elements of nonlocal quark-gluon operators. The final sum rule for  $T_1^{B^* \to K}$ has the form:
\be
2\,T_1^{B^* \to K}(q^2)\,f_{B^*}m_{B^*}\,e^{-{m_{B^*}^2 \over M^2}}= {\hat \Pi}^{QCD(0)}+{\hat \Pi}^{QCD(1)} \,\,\, ,
\label{srT1B*K}
\ee
where the symbol $\hat \Pi$  indicates that Borel transformation  and the continuum subtraction have been performed. ${\hat \Pi}^{QCD(0)}$ gets contribution only from two-particle distribution amplitudes, while  ${\hat \Pi}^{QCD(1)}$ is written in terms of the three-particle ones, all collected in the Appendix. Their  expressions are:
\bea
{\hat \Pi}^{QCD(0)}&=& f_K \, \int_{u_0}^1 {du \over u}\,e^{-{m_b^2-(1-u)q^2 \over u M^2}} \left[m_b \phi_K(u)+{m_K^2 \over m_s+m_q}\left(u\phi_P(u)+{1 \over 6} \phi_\sigma(u) \right) \right] \nonumber \\
&&+m_bf_K \left[ {1 \over M^2}\int_{u_0}^1 {du \over u}\,e^{-{m_b^2-(1-u)q^2 \over u M^2}}\, \Psi_{4K}(u) \,+{e^{-{s_0 \over M^2}} \over (s_0-q^2)}  \Psi_{4K}(u_0) \right]  \\
&&-{m_b^3 \over 4}f_K\left[{1 \over M^4} \int_{u_0}^1 {du \over u^3} \phi_{4K}(u)\,e^{-{m_b^2-(1-u)q^2 \over u M^2}}+ {e^{-{s_0 \over M^2}} \over (m_b^2-q^2)^2} \Big[\phi_{4K}(u_0) \left(1+{s_0 -q^2 \over M^2} \right)-u_0 \phi_{4K}^\prime(u_0)  \Big]\right] \,\,, \nonumber
\label{piqcd0} \eea
\bea
{\hat \Pi}^{QCD(1)}&=& \int_0^{u_0} d\alpha_1 \int_{u_0-\alpha_1}^{1-\alpha_1} d\alpha_3 \int_{u_0-\alpha_1 \over \alpha_3}^1 dv\, e^{-{m_b^2-(1-U)q^2 \over U M^2}}\nonumber \\
&&\Bigg[vf_{3K}{\phi_{3K}(\alpha_1,1-\alpha_1-\alpha_3,\alpha_3) \over U^3}\left( U-{(m_b^2-q^2) \over M^2 } \right)+{m_b f_K \over U^2 M^2}{\overline \varphi}_\perp (\alpha_1,1-\alpha_1-\alpha_3,\alpha_3)\nonumber \\
&&+{2m_b f_k \over U^3 M^2} {\overline {\hat \Phi}}(\alpha_1,1-\alpha_1-\alpha_3,\alpha_3)\left( 1-{(m_b^2-q^2) \over 2U M^2 } \right)\Bigg]\nonumber \\
&&+m_b f_K {e^{-{s_0 \over M^2}} \over (m_b^2-q^2)} \left[\int_0^1{d\alpha_3 \over \alpha_3} {\overline {\hat \Phi}}(u_0,1-u_0-\alpha_3,\alpha_3)-\int_0^{u_0}{d\alpha_1 \over u_0-\alpha_1} {\overline {\hat \Phi}}(\alpha_1,1-u_0,u_0-\alpha_1)\right] \nonumber \\
&&+e^{-{s_0 \over M^2}} \int_0^{u_0}d\alpha_1 \int_{u_0-\alpha_1}^{1-\alpha_1} {d\alpha_3 \over \alpha_3} \\
&& \Bigg[-f_{3K}{u_0-\alpha_1 \over u_0 \alpha_3}\phi_{3K}(\alpha_1,1-\alpha_1-\alpha_3,\alpha_3)
+{m_b f_K \over m_b^2-q^2}{\overline \varphi}_\perp (\alpha_1,1-\alpha_1-\alpha_3,\alpha_3)
\nonumber \\
&&-{m_b f_k \over u_0^2 M^2}{\overline {\hat \Phi}}(\alpha_1,1-\alpha_1-\alpha_3,\alpha_3)\Bigg]\nonumber \\
&&+{2m_b f_k \over M^2} \int_{1-u_0}^1 d\alpha_3 \int_0^{1-u_0 \over \alpha_3}dv \,v\, e^{-{m_b^2-(1-w)q^2 \over w M^2}} {{\overline {\hat \Psi}}(\alpha_3) \over w^3}\left(1-{m_b^2-q^2 \over 2wM^2} \right) \nonumber\\
&&+m_bf_K e^{-{s_0 \over M^2}} \left[ {{\overline {\hat \Psi}}(1-u_0) \over u_0(1-u_0)}-\int_{1-u_0}^1 {d \alpha_3 \over \alpha_3^2} {\overline {\hat \Psi}}(\alpha_3) \left({1 \over m_b^2-q^2}+{1-u_0 \over u_0^2 M^2} \right) \right] \,\,\, . \nonumber \label{piqcd1}
\eea
In  the previous equations  we have defined $u_0=\displaystyle{m_b^2-q^2 \over s_0-q^2}$, $U=\alpha_1 + v \,\alpha_3$, $w=1-v\,\alpha_3$.
Furthermore, the LCDAs have been combined as follows:
\bea
{\hat \Phi}(\alpha_1,1-\alpha_1-\alpha_3,\alpha_3) &=& -\int_0^{\alpha_1} dt \, \big[ \varphi_\perp (t,1-t-\alpha_3,\alpha_3)+\varphi_\parallel (t,1-t-\alpha_3,\alpha_3) \big]\nonumber \\
{\hat \Psi}(\alpha_3)&=&-\int_0^{\alpha_3} dt \, {\hat \Phi}(1-t,0,t)\nonumber \\
{\hat {\tilde \Phi}}(\alpha_1,1-\alpha_1-\alpha_3,\alpha_3)&=&-\int_0^{\alpha_1} dt \,\big[ {\tilde \varphi}_\perp (t,1-t-\alpha_3,\alpha_3)+{\tilde \varphi}_\parallel (t,1-t-\alpha_3,\alpha_3) \big]\nonumber \\
{\hat {\tilde \Psi}}(\alpha_3)&=&-\int_0^{\alpha_3} dt \, {\hat {\tilde \Phi}}(1-t,0,t) \\
{\overline \varphi}_\perp (\alpha_1,1-\alpha_1-\alpha_3,\alpha_3)&=& \varphi_\perp (\alpha_1,1-\alpha_1-\alpha_3,\alpha_3)-{\tilde \varphi}_\perp (\alpha_1,1-\alpha_1-\alpha_3,\alpha_3) \nonumber \\
{\overline {\hat \Phi}}(\alpha_1,1-\alpha_1-\alpha_3,\alpha_3)&=&{\hat \Phi}(\alpha_1,1-\alpha_1-\alpha_3,\alpha_3)-{\hat {\tilde \Phi}}(\alpha_1,1-\alpha_1-\alpha_3,\alpha_3) \nonumber \\
{\overline {\hat \Psi}}(\alpha_3)&=&{\hat \Psi}(\alpha_3)-{\hat {\tilde \Psi}}(\alpha_3)\,\,. \nonumber\label{definitions}
\eea
%%%%%%%%%%%%%%%%%%%%%%%%%%%%%%%%%%%%%%
The distribution amplitudes entering in the previous relations can be classified according to the twist:
$\phi_K$ is a distribution amplitude of twist 2;  $\phi_P$, $\phi_\sigma$ and $\phi_{3K}$ are twist 3; and  $\phi_{4K}$, $\psi_{4K}$, $\varphi_\parallel$, $\varphi_\perp$,  ${\tilde \varphi}_\parallel$, ${\tilde \varphi}_\perp$ are twist four.
We have used the definitions of various matrix elements  defining the LCDAs as well as the updated numerical values for their parameters in Ref.\cite{Ball:2006wn}. We use the value $M_{B^*}=5.325$ GeV, and the quark mass $m_b=4.8$ GeV which also enters in the calculation of the SM effective Hamiltonian in the next section.

Equations (\ref{srT1B*K}),(\ref{piqcd0}) and (\ref{piqcd1}) allow us to compute $T_1^{B^* \to K}$ once the threshold $s_0$ and the Borel parameter $M^2$ are fixed.
The threshold is  set to $s_0=(33\pm 2)\, {\rm GeV}^2$, a value appearing also in QCD sum rules involving the $B^*$ meson  \cite{Khodjamirian:1999hb}, which is close to the estimated mass squared of the first radial excitation of $B^*$.
 For each value of squared momentum transfer $q^2$,  the form factor $T_1^{B^* \to K}$ also depends  of the Borel parameter $M^2$, which can be fixed
 requiring stability against variations of $M^2$.
In Fig.~\ref{fig:M2dependence}  we depict the dependence of the $T_1^{B^* \to K}(0)$  on the Borel parameter $M^2$. The band reflects the uncertainties on the other quantities  entering  in the calculation,    including the uncertainties on the LCDAs parameters quoted in \cite{Ball:2006wn}, on the threshold $s_0$  and on $f_{B^*}$ for which we use the value computed by QCD sum rules in \cite{Khodjamirian:1999hb}: $f_{B^*}=0.195 \pm 0.035 $ GeV.
Although the results are quite stable with $M^2$, in the numerical analysis we fix the  stability window in the range $M^2=(17 \pm 3)\,  {\rm GeV}^2$.
%%%%%%%%%%%%%%%%%%%%%%%%%%%%%%%%%%%%%%%%%%%%%%%%%%%%%
\begin{figure}[htth]
\includegraphics[scale=0.5]{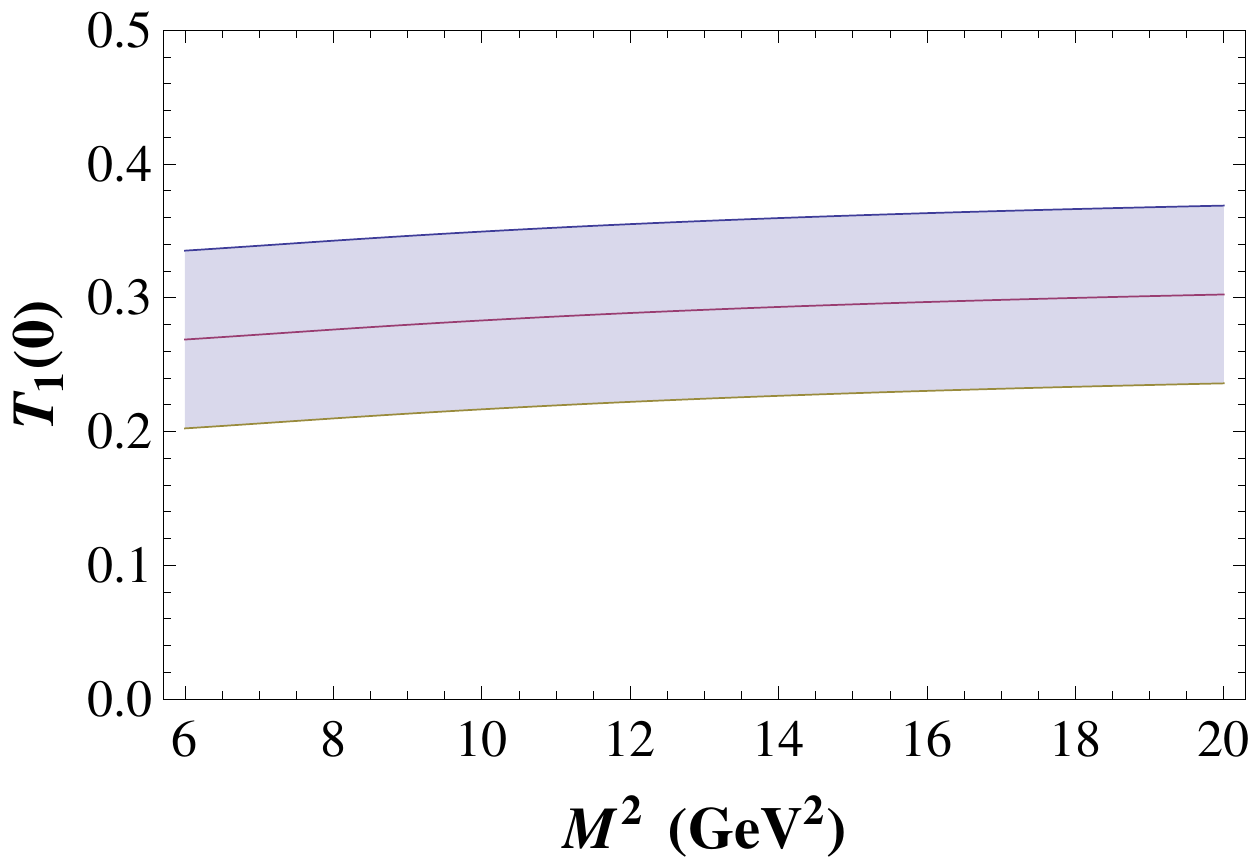}%
 \caption{Dependence  of $T_1^{B^* \to K}(0)$ on the Borel parameter $M^2$.}\label{fig:M2dependence}
\end{figure}
%%%%%%%%%%%%%%%%%%%%%%%%%%%%%%%%%%%%%%%%%%%%%%%%%%%%%

For all values of $q^2$ in the range $[0, 20]$ GeV$^2$ the computed form factor is plotted in Fig.~\ref{q2dependence}.
The functional $q^2$ dependence is obtained fitting the sum rule result by  a single  pole parametrization:
\be
T_1^{B^* \to K}(q^2)={T_1^{B^* \to K}(0) \over 1- \displaystyle{q^2 \over M_P^2}} \label{t1fit}
\ee
with
\be
T_1^{B^* \to K}(0)=0.30 \pm 0.066 \hskip 2 cm M_P=5.767 \, {\rm GeV} \,\,\,.
\label{par-fit}
\ee
The uncertainty of  $T_1^{B^* \to K}(0)$  also  accounts for the variation of the Borel parameter within the stability window.

%%%%%%%%%%%%%%%%%%%%%%%%%%%%%%%%%%%%%%%%%%%%%%%%%%%%%
\begin{figure}[htth]
\includegraphics[scale=0.5]{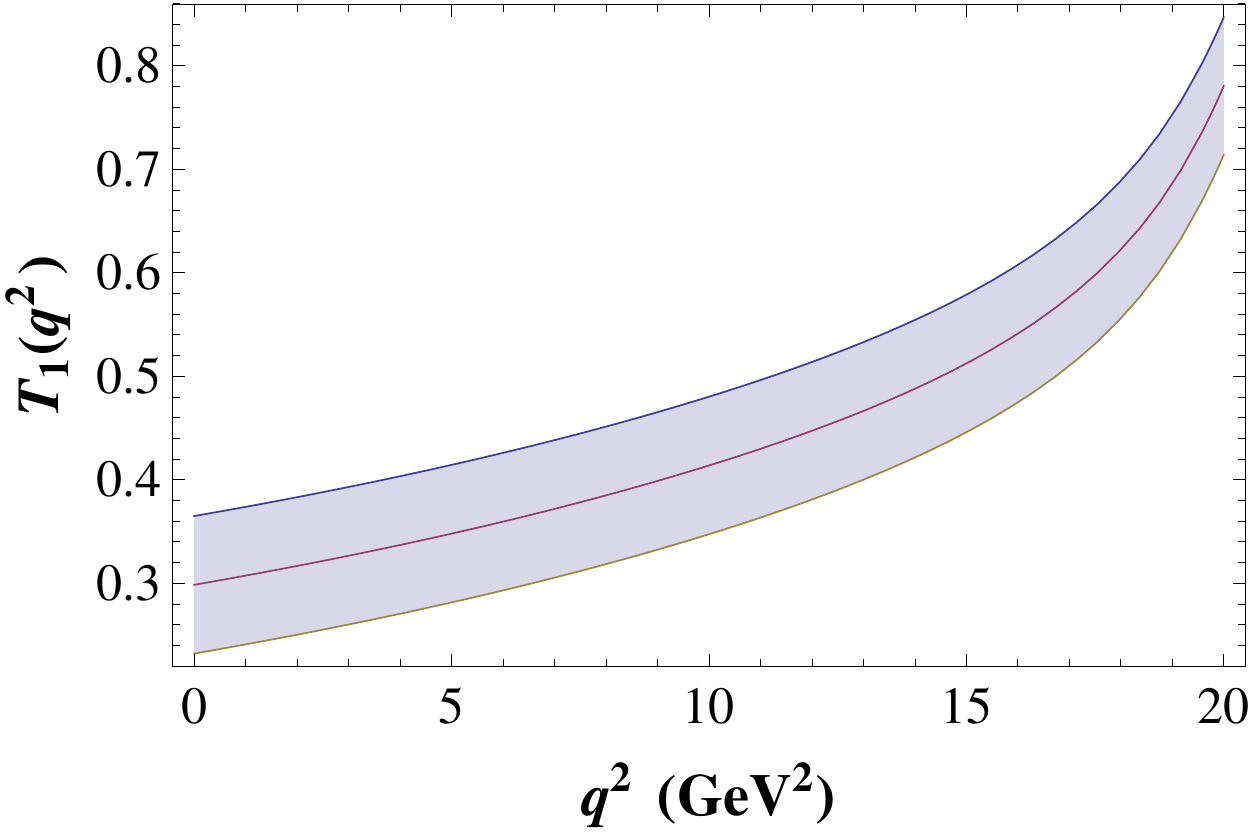}%
 \caption{ $q^2$ dependence of the form factor $T_1^{B^* \to K}(q^2)$.}\label{q2dependence}
\end{figure}
%%%%%%%%%%%%%%%%%%%%%%%%%%%%%%%%%%%%%%%%%%%%%%%%%%%%%

We conclude this section with a comment about the relations among the form factors that parametrize the $B$ to a light  meson $L$ matrix elements.  As shown in Ref.\cite{Charles:1998dr},   when the energy of $L$
in the rest-frame of the decaying $B$ meson  is large,  the form factors describing the  $B$ to $L$ transitions can be related among each other.
Considering also the heavy quark limit, it is possible to relate the $B \to L$  form factors to the $B^* \to L$ ones, as shown in \cite{Beneke:2000wa} where also the perturbative corrections to the large energy relations have been worked out.
In particular, the relation  $T_1^{B^* \to K}=f_+^{B \to K}$ should hold for large  kaon energy, where $f_+$ parametrizes the matrix element of the quark vector current ${\bar s} \gamma_\mu b$ between the kaon and the $B$ meson as in the first equation in (\ref{bkphi-fact}). Since large kaon energy means  $q^2$ close to $q^2=0$, one might exploit the relation $T_1^{B^* \to K}(0)=f_+^{B \to K}(0)$, as done in \cite{Fajfer:2008zy}.
Our  computed form factor $T_1^{B^* \to K}$, together with    $f_+^{B \to K}(0)=0.331 \pm 0.041$ determined by LCSR \cite{Ball:2004ye},  fulfills the relation  within the uncertainties.

\section{$B \to K \eta^{(\prime)} \gamma$ decay rates and photon spectra}\label{numerics}
In Secs.\ref{calc} and \ref{LCSR} we have  collected  the quantities necessary to evaluate the amplitudes corresponding to the diagrams in Fig. \ref{diagrams}. Input parameters are  the quark masses, $m_b$ already fixed  and $m_s=0.130$ GeV \cite{Colangelo:2000dp},  and the CKM matrix elements $V_{tb}=0.99$, $V_{ts}=0.04$ \cite{pdg}. We do not include the uncertainties on these parameters because they are small with respect to
the uncertainties of the other input quantities; moreover, $m_s$ plays a negligible role in the final  result. We set the renormalization scale at which we compute the  coefficient $C_{7\gamma}^{(eff)}$  to $\mu=5$ GeV,
close to the $b$ mass.

We can now  discuss the branching fraction and the photon spectrum of  $B \to K \eta \gamma$ and $B \to K \eta^\prime \gamma$
 in the SM and in two new physics models with universal extra dimensions described  below. We anticipate that
 such new physics scenarios belong to the class of minimal flavor violation models, therefore the only modification  with respect to  the SM consists in a different value of the Wilson coefficients in the effective weak Hamiltonian.
Therefore, the three cases, the SM and the two UEDs,  share  common features, namely  the hierarchy among the various decay amplitudes  and the shape of the photon spectrum.

For what concerns  the intermediate states  in the $B \to K \eta^{(\prime)} \gamma$ decay amplitudes,   the most important contributions are represented by the diagrams (1), (3) and (4) in the case of the $\eta$, while for $\eta^\prime$ the first diagram contributes much less than diagrams (3) and (4).
This is due to the coupling $g_{K^*K\eta}$ which is much larger than $g_{K^*K\eta^\prime}$: indeed, from the relations in Sec.\ref{calc} we get $g_{K^*K\eta}=11 \pm 0.1$ and $g_{K^*K\eta^\prime}=-2.57 \pm 0.19$.
 We only consider a phase $\theta$ between the sum of the first two amplitudes and  $A_3+A_4$;  the fifth diagram turns out to be much smaller than the others, hence   we assign to it the same phase as  $A_3$ and $A_4$ since a wide change of its phase does not modify the result. In the case of $\eta^\prime$ we do not consider the
contribution of $K^*_2$, and $\theta$ is the phase between $A_1$ and $A_3+A_4+A_5$. From the calculation of   ${\cal B}(B \to K \eta \gamma)$  we shall see that there is a range of values of   $\theta$ allowing to reproduce experimental data in Table \ref{table:br}.  Let us  start from the standard model.

\subsection{$B \to K \eta^{(\prime)} \gamma$ in the standard model}

The plot of the computed  ${\cal B}(B \to K \eta \gamma)$ as a function of the strong phase $\theta$ is  depicted in  Fig. \ref{phase-eta}.  The  experimental results   in Table \ref{table:br} can be obtained  for
 $\theta=1.8 \pm 1.0$ rad, corresponding approximately to $\theta=\left(7\pm 3 \right) \frac{\pi}{12}$  (we have considered the range $0 \leq \theta \leq\pi$ since the plot is symmetric with respect to $\theta=0$; indeed, in the branching ratio the term proportional to $\sin \theta$ is 2 orders of magnitude smaller than the one proportional to $\cos \theta$).
%%%%%%%%%%%%%%%%%%%%%%%%%%%%%%%%%%%%%%%%%%%%%%%%%%%%%
\begin{figure}[h!]
\includegraphics[scale=0.5]{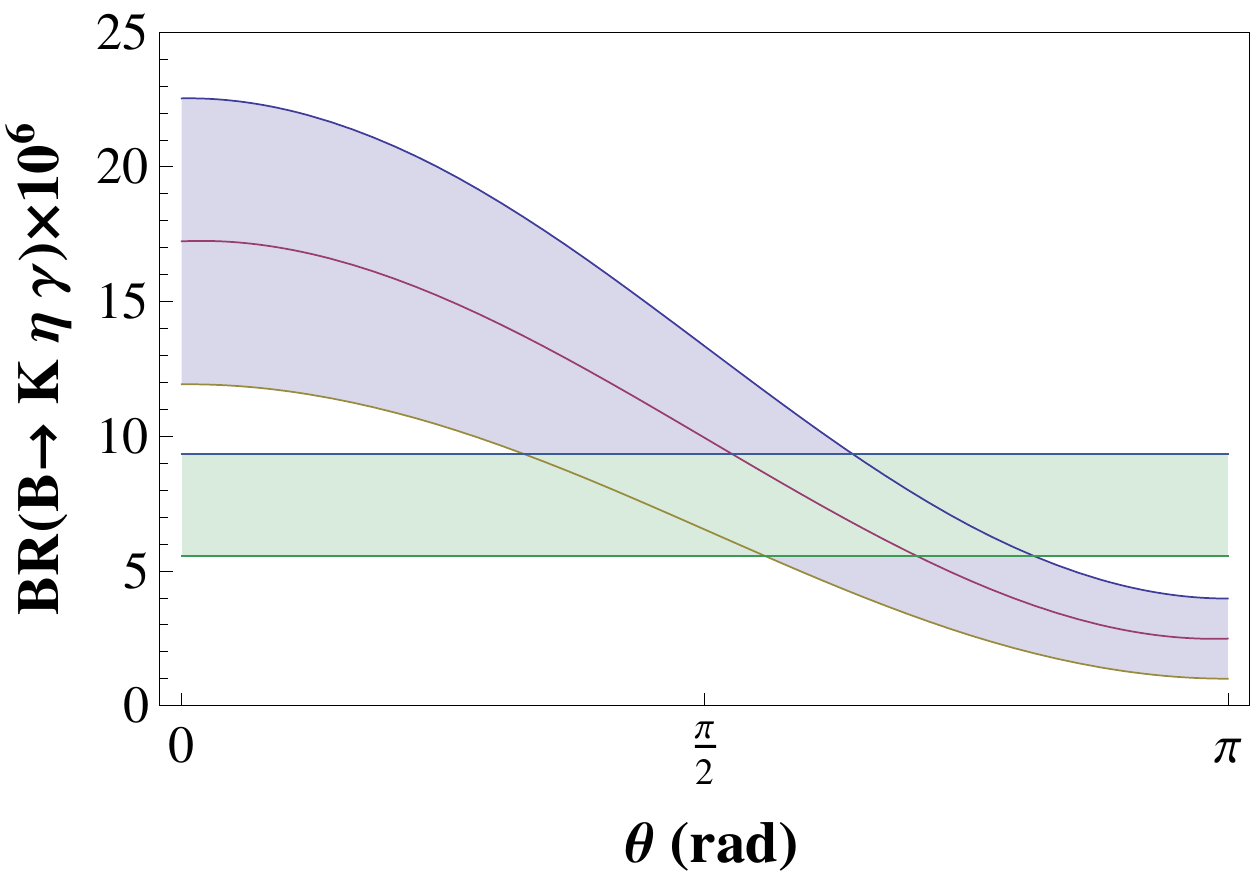}
 \caption{Branching ratio ${\cal B}({\overline B^0} \to {\overline K^0} \eta \gamma)$ as a function of the strong phase phase $\theta$ between $A_1+A_2$ and  $A_3+A_4+A_5$. The horizontal band corresponds to the experimental result. }\label{phase-eta}
\end{figure}
%%%%%%%%%%%%%%%%%%%%%%%%%%%%%%%%%%%%%%%%%%%%%%%%%%%%%
For the central value of  $\theta$,   the  photon spectrum is depicted in Fig. \ref{spettro-eta}. It is peaked at  large photon energies and  has  a structure  as the effect of the virtual $K^*$. The Dalitz plot in the plane $( M_{\eta K}, \, E_\eta)$,  displayed in Fig. \ref{Dalitzeta}, also shows the effect of the  $K^*$ in $B \to K \eta \gamma$,
 at the limit of the  phase space:  it  should be observed in the data.

 In the  case of the $\eta^\prime$, since the diagram (1) gives a small contribution with respect to (3) and (4),   strong phases do not play any role. The result for the branching ratio is ${\cal B}(B \to K \eta^\prime \gamma)=(2.78\pm 1.14)\times 10^{-7}$, with
the photon spectrum  depicted in Fig.\ref{spettro-eta} and the Dalitz plot  shown in Fig.\ref{Dalitzeta}.
The theoretical result  of the branching fraction for the neutral mode is compatible with the  experimental datum  in Table \ref{table:br}, which is affected by a  large uncertainty.
The experimental error is smaller in the charged mode: in this case, while the BaBar result is compatible with the calculation,  the Belle measurement is larger. Before commenting on the charged mode,
it is worth observing that, for the $B \to K \eta^{(\prime)} \gamma$ three-body channels,  the hierarchy between the modes with  $\eta$ and   $\eta^\prime$, observed in data,  is reproduced by the theoretical calculation in the frameworks of the QF scheme for the $\eta-\eta^\prime$ mixing.

%%%%%%%%%%%%%%%%%%%%%%%%%%%%%%%%%%%%%%%%%%%%%%%%%%%%%
\begin{figure}[htth]
\includegraphics[scale=0.5]{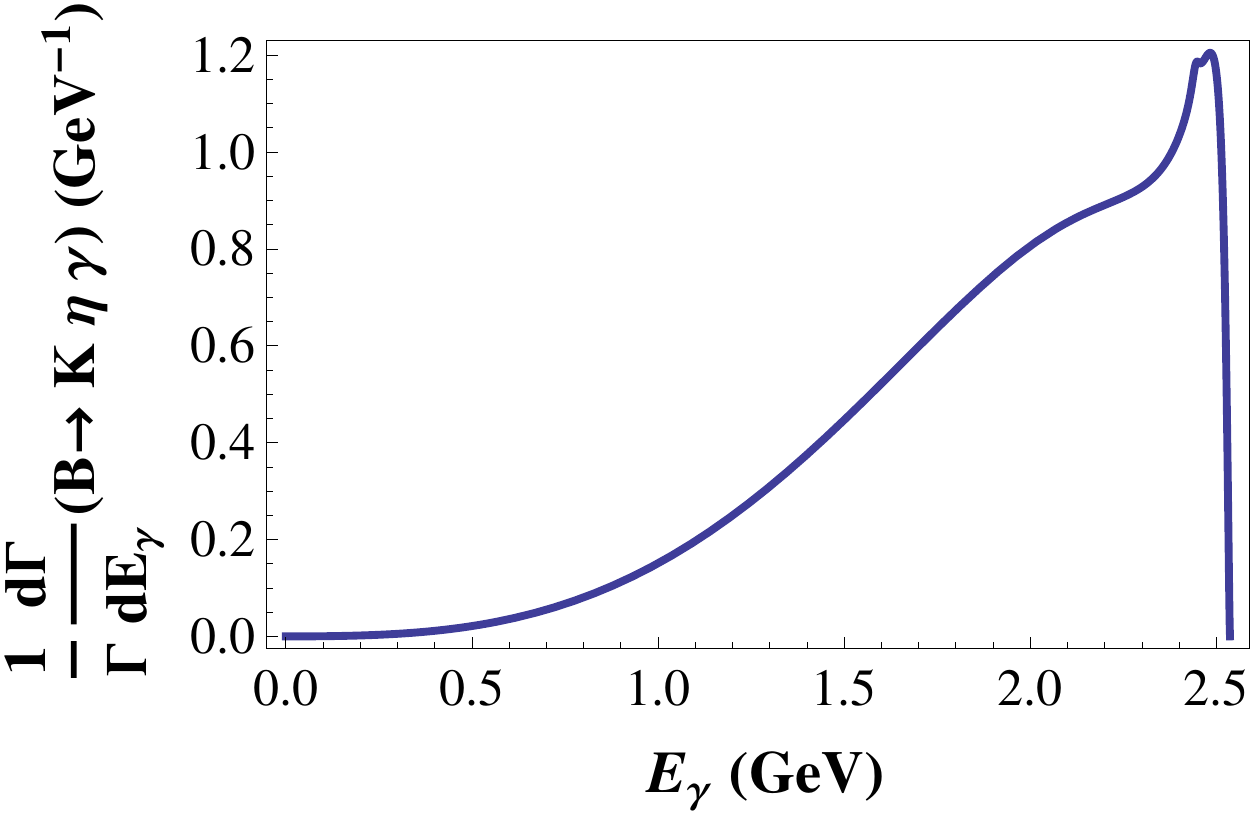}\hspace*{1.5cm}
\includegraphics[scale=0.5]{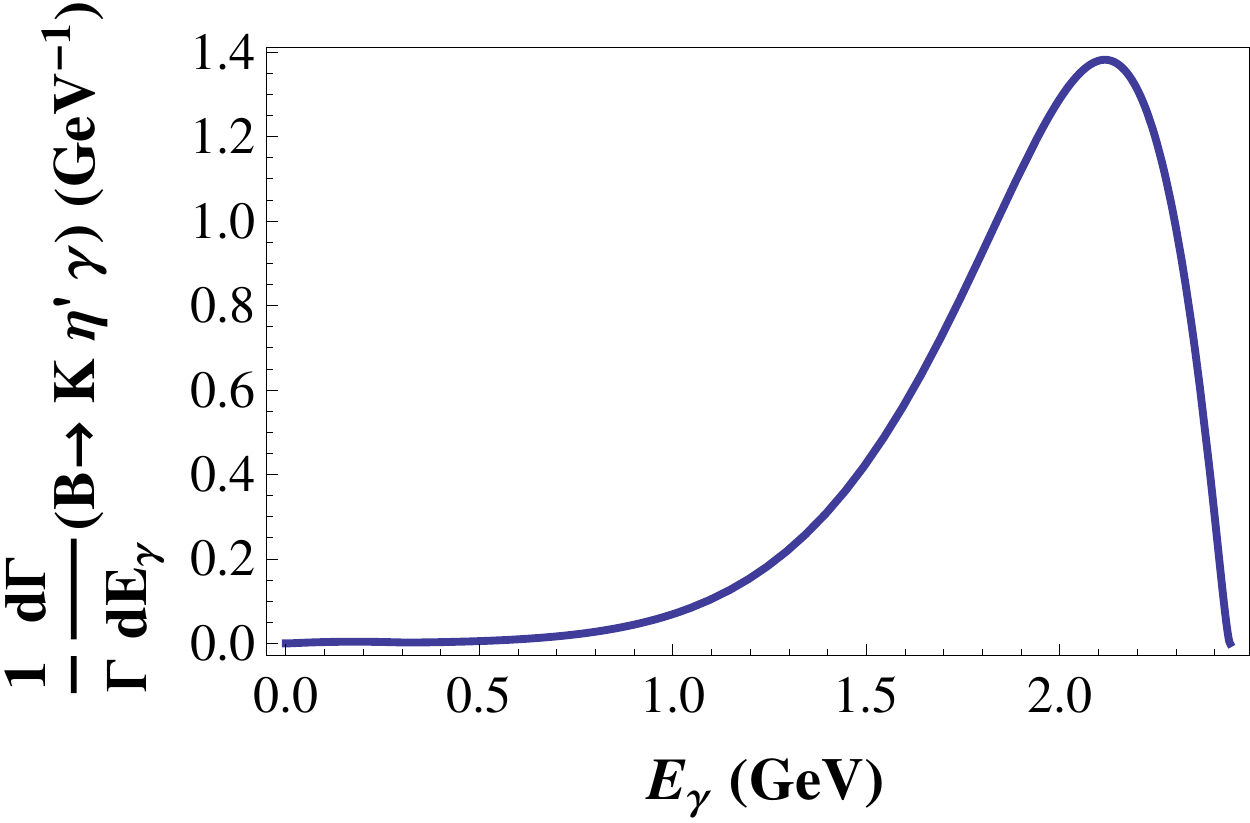}
 \caption{Photon spectrum in  ${\overline B^0} \to {\overline K^0}  \eta \gamma$ (left) and ${\overline B^0} \to {\overline K^0} \eta^\prime \gamma$ (right). The  phase $\theta$ is set to $\theta=1.8$ rad.}\label{spettro-eta}
\end{figure}
%%%%%%%%%%%%%%%%%%%%%%%%%%%%%%%%%%%%%%%%%%%%%%%%%%%%%

Let us discuss the differences between the neutral ${\overline B^0}$ and the charged $B^{\pm}$ radiative decays.  The  analysis of the charged modes would be the same as the one
we have presented for the neutral modes, except for  the  contribution of  the inner bremsstrahlung  diagrams  with the photon coupled to the charged initial $B^+$ and final $K^+$ mesons. The  kinematical region in which the bremsstrahlung could be competitive with the other decay mechanisms is for soft photons, due to the presence of a pole at vanishing  photon energy.
To describe this contribution to the $B^+ \to K^+ \eta^{(\prime)} \gamma$ decay, we invoke the Low theorem \cite{Low:1958sn} which, for scalar particles,   allows to relate the amplitude of the radiative mode with a soft photon  to the amplitude  $A(B^+ \to K^+ \eta^{(\prime)})$:
\be
A_{IB}(B^+(p) \to K^+(p_K) \eta^{(\prime)}(p_{\eta^{(\prime)}}) \gamma (q,\epsilon))=e\,\left( {\epsilon^* \cdot p_K \over q \cdot p_K} -  {\epsilon^* \cdot p_{\eta^{(\prime)}} \over q \cdot p_{\eta^{(\prime)}}} \right)\,A(B^+ \to K^+ \eta^{(\prime)})\,\,. \label{low}
\ee
The two-body amplitudes in (\ref{low}) can be   determined from the experimental branching fractions  ${\cal B}(B^+ \to K^+ \eta)=(2.33\pm^{0.33}_{0.29})\times 10^{-6}$ and ${\cal B}(B^+ \to K^+ \eta^\prime)=(7.06\pm 0.25 )\times10^{-5}$ \cite{pdg}.  In the case of  the $\eta$,   the estimated  contribution of the inner bremsstrahlung diagram to the decay rate is of order ${\cal O}(10^{-8})$;
therefore, the rate of the charged mode is not  significantly affected by this effect,
as indeed observed in the data in Table \ref{table:br}.
The contribution is more important for the $\eta^\prime$: varying  the relative phase $\theta_{brem}$ between the bremsstrahlung amplitude and the other considered amplitudes, ${\cal B}(B^+ \to K^+ \eta^\prime \gamma)$ varies between $(2.3 \pm 1.0) \times 10^{-7}$  and $(3.8 \pm 1.4) \times 10^{-7}$. These results are within $1 \sigma$ from the central value of the BaBar measurement in Table \ref{table:br}, while they deviate by about $2.5 \sigma$ from the Belle data.
We do not further elaborate on this point: if the deviation is confirmed (or strengthened) by new measurements, the interesting issue of additional contributions to the $B^+ \to K^+ \eta^\prime \gamma$ amplitude must be addressed.

\vspace*{0.5cm}
\begin{figure}[h!]
\begin{center}
\includegraphics[scale=0.5]{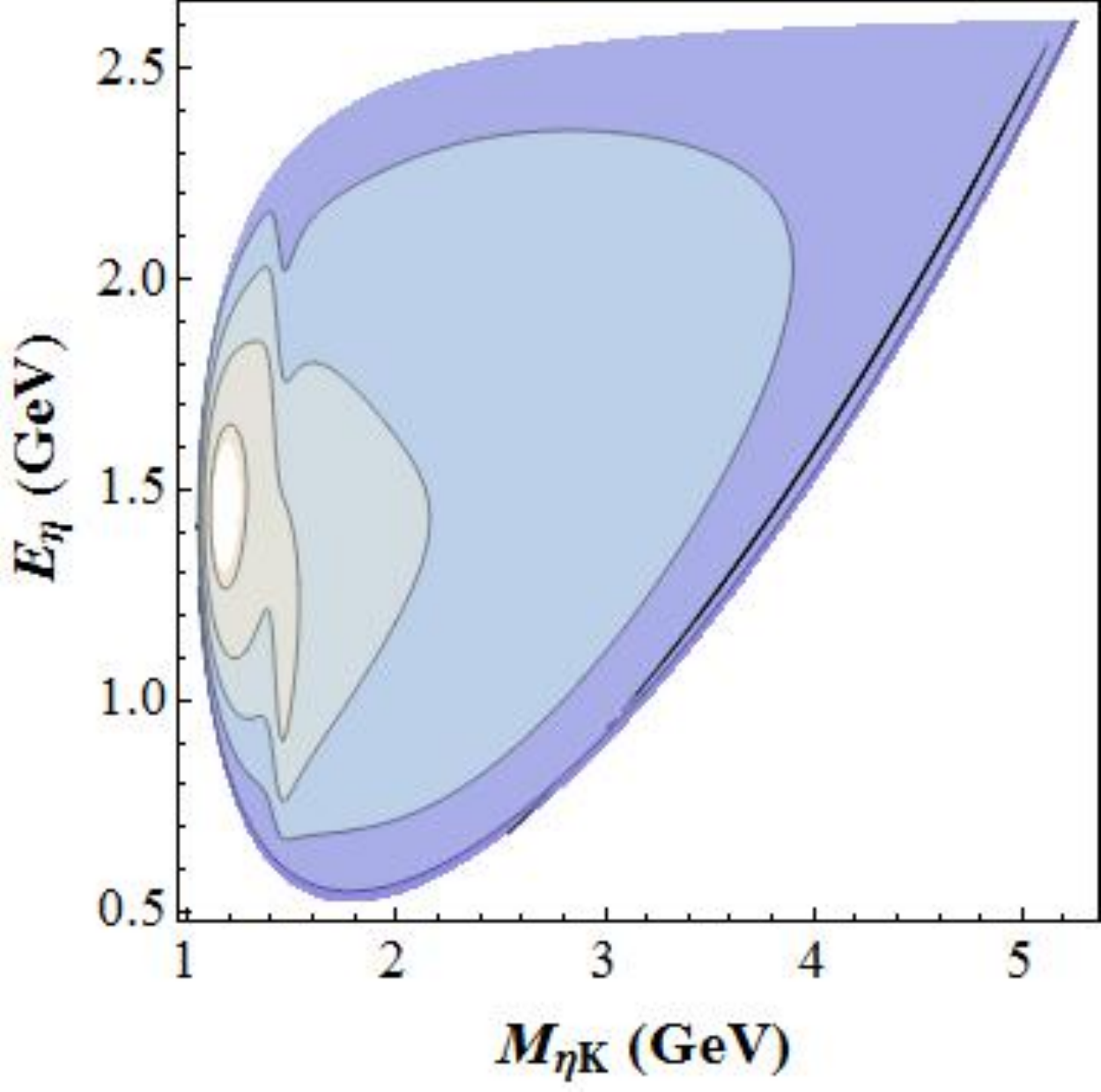}\hspace{1.5cm}
\includegraphics[scale=0.5]{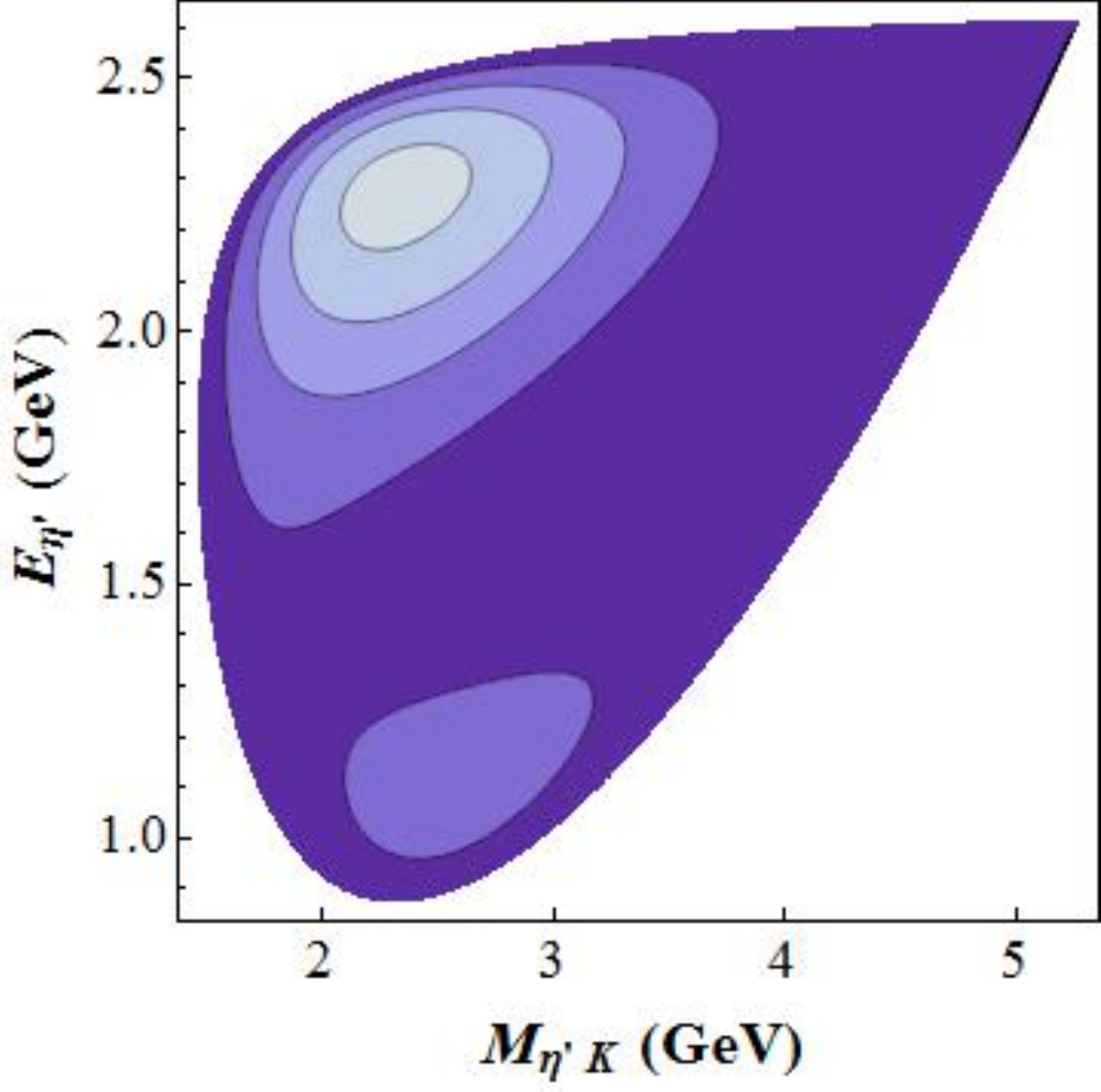}
\end{center}
 \caption{Dalitz plot of ${\overline B^0} \to {\overline K^0}  \eta \gamma$  (left panel) and ${\overline B^0} \to {\overline K^0} \eta^\prime \gamma$  (right panel)
  in the plane $(M_{\eta^{(\prime)} K}, E_{\eta^{(\prime)}})$. The photon and the $\eta^{(\prime)}$ energy in the $B$ rest frame  are
 $\displaystyle E_\gamma=\left({M^2_B-M^2_{\eta^{(\prime)} K}}\right)/{2 M_B}$ and  $\displaystyle E_{\eta^{(\prime)}}=\left({M^2_B+M^2_{\eta^{(\prime)} }-t}\right)/{2 M_B}$, respectively.
Lighter colors correspond to higher values of the distribution.}\label{Dalitzeta}
\end{figure}

\subsection{Sensitivity of $B \to K \eta \gamma$ to two  new physics  UED scenarios}
It is worth investigating the sensitivity of the rare FCNC  $B \to K \eta \gamma$ transition to new physics effects. In particular, it is important to establish which kind of improvement can be achieved by a more precise determination of the branching fraction.
The  considered new physics scenarios involve one or two universal extra dimensions (UED).

The  scenario with a  single universal extradimension is  the Appelquist-Cheng-Dobrescu (ACD) model \cite{Appelquist:2000nn} \footnote{One of the first proposals to introduce large (TeV) extra dimensions in the SM was suggested in \cite{Antoniadis:1990ew}.},  a minimal extension of
SM in $4+1$ dimensions,  with the extra dimension compactified to
the orbifold $S^1/Z_2$ and the fifth coordinate $y$ running from $0$ to
$2 \pi R$,   $y=0$ and $y=\pi R$  being fixed points of the orbifold. All the fields   propagate
 in all  $4+1$ dimensions,  therefore the model belongs to the class of  {\it universal} extra dimension scenarios;  one of its  motivations
is that it naturally provides candidates for the dark matter.

In the ACD model  the SM particles correspond to  the zero modes of fields propagating in the
compactified extra dimension. In addition to the zero modes,    towers of
Kaluza-Klein (KK) excitations are predicted to exist, corresponding to the higher modes of the fields in the extra dimension;
such fields  are imposed to be even under a   $y \to -y$ transformation in the fifth coordinate.  On the other hand,  fields  which are odd under this transformation propagate in the extra
dimension without zero modes, and correspond to particles  without  SM partners.

The masses of  KK particles  depend on  the radius $R$ of the compactified extra dimension,  the  new parameter with respect to SM.
For example,
  the  masses of the KK bosonic modes are given by
 \be m_n^2=m_0^2+{n^2 \over R^2} \,\,\,\,\,\,\, n=1,2,\dots \ee
 $m_0$ being the mass of the zero mode, so that   for small values of $R$, i.e. at large compactification scales,  the KK particles decouple from the low-energy regime.
 Another  property of the ACD model is  the conservation of the KK parity $(-1)^j$,  $j$ being  the KK number.  KK parity
conservation implies  the absence of tree level contributions of
Kaluza-Klein states to processes taking place at low energy,    forbidding the production of a single KK particle  off the interaction of standard particles.  This permits to use the electroweak measurements  to provide a lower bound to the compactification scale:
   ${1 / R} \ge 250-300$ GeV \cite{Appelquist:2002wb}.
 Moreover, this suggests  the possibility that the lightest  KK particles,  namely the $n=1$
Kaluza-Klein excitations of the photon and  neutrinos, are among  the  dark matter components  \cite{Cheng:2002iz,Hooper:2007qk}.

Since KK modes  affect the loop-induced processes,  flavor changing neutral current  transitions can constrain this new physics scenario.   Indeed,  many observables are sensitive to the compactification radius in the case, e.g., of processes involving $B$, $B_s$ and $\Lambda_b$
  \cite{buras,noi0,noi,UEDvarie,UEDvarie1,UEDvarie2}.
  In the ACD model  no operators other than those in (\ref{eff}) contribute to  $b \to s \gamma$, and the  effects beyond SM are only encoded in the Wilson coefficients of the effective Hamiltonian.
 The contribution of  KK excitations   modifies in particular  the  coefficient $C_{7\gamma}$,  which  acquires a
dependence on the  compactification scale $1/R$.  For large values
of $1/R$, due to decoupling of  massive KK states,    the   coefficient  $C_{7\gamma}$ (whose explicit expression  can be found in  \cite{buras})  reproduces the  standard model value.
For this scenario, the bound $\displaystyle \frac{1}{R}>600$ GeV has been established from exclusive \cite{noi0} and inclusive \cite{UEDvarie} radiative rare $B$ decays.

%%%%%%%%%%%%%%%%%%%%%%%%%%%%%%%%%%%%%%%%%%%%%%%%%%%%%
\begin{figure}[b!]
\subfigure[]{\includegraphics[scale=0.565]{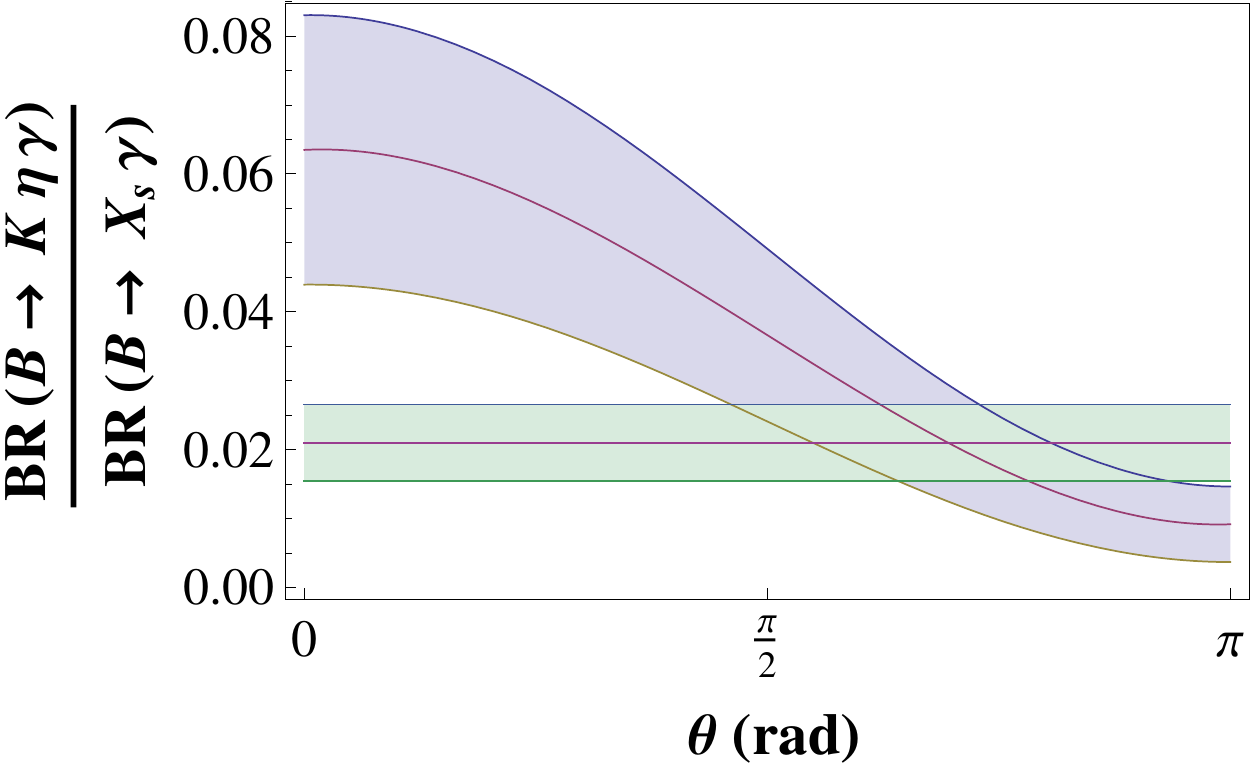}\hspace{1.cm}   \label{eta-acd}}
\subfigure[]{\includegraphics[scale=0.5]{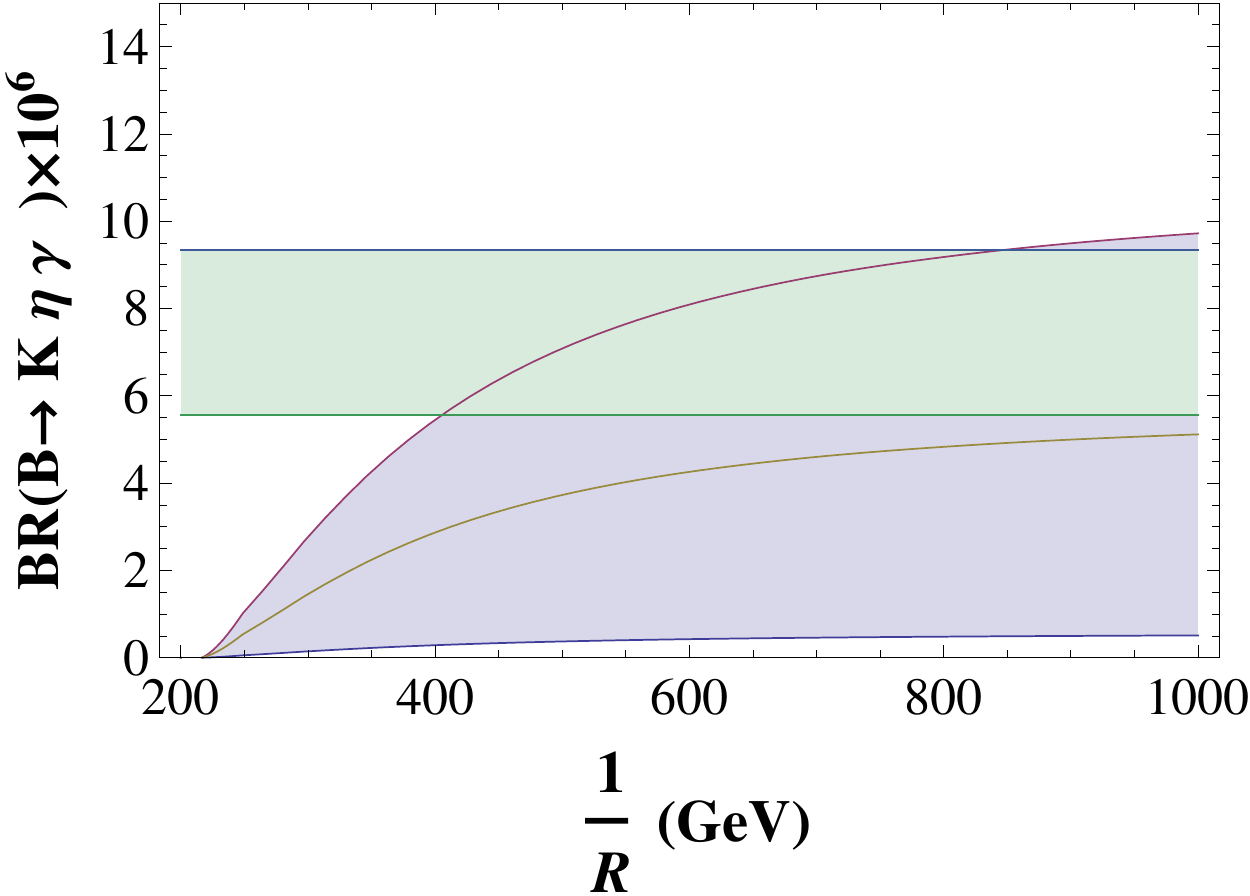}\label{eta_acd2}}
 \caption{Ratio of the experimental branching fractions
 $\displaystyle \frac{ {\cal B}({\overline B^0} \to {\overline K^0} \eta \gamma)}{{\cal B}({\overline B^0} \to X_s \gamma)}$ as a function of the phase $\theta$ (left panel),  and
 branching fraction ${\cal B}({\overline B^0} \to {\overline K^0} \eta \gamma)$ computed in  the model with  two universal extra dimensions as a function
 of the inverse of the compactification radius (in GeV) and for the phase in the range fixed in (a) (right panel).
 The horizontal bands correspond to the experimental data.}\label{bracd-eta}
\end{figure}
%%%%%%%%%%%%%%%%%%%%%%%%%%%%%%%%%%%%%%%%%%%%%%%%%%%%%

The second scenario we consider involves  two   UEDs \cite{Dobrescu:2004zi}. In this case,
the two extra dimensions are flat and compactified on a (so-called chiral) square of side $L$: $0 \le x^4,x^5 \le L$, where $x^4$ and $x^5$ are  the fifth and sixth extra spatial coordinates. The compactification is performed identifying two pairs of adjacent sides of the square: $(y,0)=(0,y)$ and $(y,L)=(L,y)$,  for all $y \in [0,L]$,  which amounts to folding the square along a diagonal. The fields are decomposed in Fourier modes in terms of effective four dimensional fields labeled by two indices $(l,k)$. Hence,  the KK modes are identified by two KK numbers which determine their mass in four dimensions:  zero modes corresponds to SM fields.
The values of the fields in the points identified through the folding are related by a symmetry transformation. For example, for a scalar field,  the
 field values
may differ by a phase. The choice of the folding boundary conditions (and of the constraints on such phases) is mostly important in the case of fermions, since a suitable choice allows to obtain chiral zero modes, while higher KK modes have masses determined (as for scalars) by the relation: $M_{l,k}=\displaystyle{\sqrt{l^2+k^2} \over R}$, where $R=\displaystyle{L \over \pi}$ is the compactification radius.
The theory has an additional symmetry, the  invariance under reflection with respect to the center of the square. Such a symmetry distinguishes between the various KK excitations of a given particle. A KK mode identified by the pair $(l,k)$ of indices changes sign under reflection if $l+k$ is odd, while it remains invariant if $l+k$ is even. As a consequence, the stability of the lightest KK modes is guaranteed, and  such  modes are good candidates  for dark matter.

The model comprises  the SM particles and their KK excitations,  together with new particles without a SM correspondent,   described by fields whose Fourier decomposition does not contain a zero mode. Examples are the mixing of the fifth and fourth components of the vector fields and the Higgs fields. All these new particles may contribute as intermediate states in the FCNC loop diagrams and, as in the single UED case, they modify the values of the Wilson coefficient in the effective Hamiltonian (\ref{eff}) without introducing  new operators. The explicit expression of $C_{7\gamma}^{eff}$ in this model can be found  in Ref.\cite{Freitas:2008vh}. It should only be mentioned that the sums over the KK modes entering in the expression of the Wilson coefficients in the extra dimensional framework diverge logarithmically, and should be
 cut in correspondence of some values of $N_{KK}=l+k$, viewing  this theory  as an effective one valid up to a some higher scale. The  condition  $N_{KK} \simeq 10$ has been chosen in  \cite{Freitas:2008vh}.

In order to disentangle the dependence of the rate $B \to K \eta\gamma$ on the phase $\theta$ and on the Wilson coefficient $C_7$ which encodes the new physics effects, we  consider the ratio
$\displaystyle \frac{BR({\overline B^0} \to {\overline K^0} \eta\gamma)}{BR({\overline B^0} \to X_s \gamma)}$  versus  $\theta$, with the experimental datum of $BR({\overline B^0} \to X_s \gamma)$ reported in \cite{hfag} and the theoretical expression that can be found, e.g. in \cite{Ewerth:2009yr}. In this ratio, the dependence on $C_7$ cancels out, so that we can fix the range of allowed values of the phase depending on the
experimental measurements with their own accuracy. As depicted in  Fig. \ref{eta-acd}, the data allow to determine a range for  $\theta$ (a strong interaction quantity): $\theta=2.19\pm0.75$ rad, which is compatible with the
range determined in the previous section and
 can be  reduced by improved measurements of the decay rates. With $\theta$ in this range
and the expression  of $C_{7\gamma}^{eff}$  dependent in both the models  on the respective compactification radii, we can compute    $BR({\overline B^0} \to {\overline K^0} \eta\gamma)$  versus $\displaystyle 1/R$. While for the case of the ACD model no sensible bound
on $1/R$ can be worked out, in the case of two UEDs, as
plotted in Fig. \ref{eta_acd2},   the constraint $\displaystyle{1 \over R}>400$ GeV can be derived. Although such a constraint is weaker than the  bound established from  the inclusive radiative rare $B$ decay rate,
$\displaystyle{1 \over R}>650$ GeV  \cite{Freitas:2008vh},
it represents an additional information that can be made more precise, e.g.,  improving the experimental data.

\section{Conclusions}\label{conclusion}
$B$ decays to two light pseudoscalar mesons and a photon are interesting,  as witnessed by the experimental efforts to determine,
 for the modes  $B \to K \eta^{(\prime)} \gamma$ considered here,   the branching fractions and the CP asymmetry parameters.
 We have studied such channels  considering the contribution of amplitudes corresponding to several  intermediate states, $K^*(892)$ and $K^*_2(1430)$, as well as  $B^*$, $B^*_s$ and $\phi(1020)$.
A  light cone sum rule  determination of
the form factor $T_1^{B^* \to K}(q^2)$ has been performed: this form factor is of interest since it also enters in other amplitudes involving  $B^*$ mesons.
 Introducing a strong  phase $\theta$  between the  first two considered contributions and  the other three, we have shown that  the measured $B \to K \eta \gamma$ branching fraction  can be reproduced. On the other hand, the experimental uncertainties in  ${\cal B}(B \to K \eta^\prime \gamma)$ are large, so that  the comparison with the theoretical result   does not provide constraints, at present.
In any case, the modes with $\eta^\prime$ in the final state are not enhanced with respect to those with the $\eta$, as experimentally observed.
The  photon spectrum, as well as the Dalitz plots, are sensitive to the intermediate contributions.

We have studied the radiative transitions in NP scenarios with  one and two universal extra dimensions, to study their sensitivity to NP effects. We have found that,  in the case of two UEDs compactified on the chiral square,
 the bound $\displaystyle{1 \over R}>400$ GeV can be established from ${\cal B}({\overline B^0} \to {\overline K^0} \eta \gamma)$.

\section*{Acknowledgement}
We  thank  A.~J.~Buras and  E.~Scrimieri for useful discussions. This work is supported in part by the Italian MIUR Prin 2009.

\appendix*
\section{LCDAs of the $K$ meson}
Here we collect the matrix element defining the LCDAs of the kaon used for the calculation of the  form factor $T_1^{B^* \to K}(q^2)$ .
\begin{itemize}
\item Two-particle LCDAs
\end{itemize}
\be
\langle K(p^\prime) |{\bar s}(x) \gamma_\mu \gamma_5 q(0)|0 \rangle=-i\,p^\prime_\mu \int_0^1 du e^{i\,u p^\prime \cdot x}[\phi_K(u)+{x^2 \over 16} \phi_{4K}(u)]-{i \over 2}f_K {x_\mu \over p^\prime \cdot x} \int_0^1
e^{i\,u p^\prime \cdot x} \psi_{4K}(u)\,\,; \nonumber
\ee
\be
\langle K(p^\prime) |{\bar s}(x) i \gamma_5 q(0)|0 \rangle= {f_K m_K^2 \over m_s +m_q} \int_0^1 du e^{i\,u p^\prime \cdot x} \phi_P(u) \,\,; \nonumber
\ee
\be
\langle K(p^\prime) |{\bar s}(x) \sigma_{\mu \nu} \gamma_5 q(0)|0 \rangle= i \,(p^\prime_\mu x_\nu-p^\prime_\nu x_\mu){f_K m_K^2 \over6( m_s +m_q)} \int_0^1 du e^{i\,u p^\prime \cdot x} \phi_\sigma
(u) \,\,. \nonumber
\ee
\begin{itemize}
\item Three-particle LCDAs
\end{itemize}
\bea
&&\langle K(p^\prime) |{\bar s}(x) g_s G^{\theta \tau} \sigma_{\rho \nu} \gamma_5 q(0)|0 \rangle= \nonumber \\
&&i \, f_{3K} [(p^{\prime\theta} p^\prime_\rho g^\tau_\nu -p^{\prime\tau} p^\prime_\rho g^\theta_\nu)-
(p^{\prime\theta} p^\prime_\nu g^\tau_\rho -p^{\prime\tau} p^\prime_\nu g^\theta_\rho)] \int {\cal D} \alpha_i \phi_{3K}(\alpha_1,\alpha_2,\alpha_3) e^{i \, p^\prime \cdot x \, (\alpha_1+v \,  \alpha_3)} \,\,; \nonumber
\eea
\bea
&& \langle K(p^\prime) |{\bar s}(x) g_s G^{\theta \tau} \gamma^\psi \gamma_5 q(0)|0 \rangle= \nonumber \\
&& f_K \left[ p^{\prime \tau} \left( g^{\theta \psi} -{x^\theta p^{\prime \psi} \over p^\prime \cdot x} \right) -
p^{\prime \theta} \left( g^{\tau \psi} -{x^\tau p^{\prime \psi} \over p^\prime \cdot x} \right) \right]
\int {\cal D} \alpha_i \, e^{i \, p^\prime \cdot x \, (\alpha_1+v \,  \alpha_3)} \, \varphi_\perp (\alpha_1,\alpha_2,\alpha_3) \nonumber \\
&&+ f_K {p^{\prime \psi} \over p^\prime \cdot x}  (p^{\prime\theta} x^\tau -p^{\prime\tau} x^\theta)
\int {\cal D} \alpha_i \, e^{i \, p^\prime \cdot x \, (\alpha_1+v \,  \alpha_3)} \, \varphi_\parallel (\alpha_1,\alpha_2,\alpha_3) \nonumber
\,\,;
\eea
\bea
&& \langle K(p^\prime) |{\bar s}(x) g_s {\tilde G}^{\theta \tau} \gamma^\psi  q(0)|0 \rangle= \nonumber \\
&& f_K \left[ p^{\prime \tau} \left( g^{\theta \psi} -{x^\theta p^{\prime \psi} \over p^\prime \cdot x} \right) -
p^{\prime \theta} \left( g^{\tau \psi} -{x^\tau p^{\prime \psi} \over p^\prime \cdot x} \right) \right]
\int {\cal D} \alpha_i \, e^{i \, p^\prime \cdot x \, (\alpha_1+v \,  \alpha_3)} \, {\tilde \varphi}_\perp (\alpha_1,\alpha_2,\alpha_3)
\nonumber \\
&&+ f_K {p^{\prime \psi} \over p^\prime \cdot x}  (p^{\prime\theta} x^\tau -p^{\prime\tau} x^\theta)
\int {\cal D} \alpha_i \, e^{i \, p^\prime \cdot x \, (\alpha_1+v \,  \alpha_3)} \, {\tilde \varphi}_\parallel (\alpha_1,\alpha_2,\alpha_3) \nonumber
\,\,.
\eea
The following definitions have been used: ${\cal D} \alpha_i=d \alpha_1 d \alpha_2 d \alpha_3 \, \delta(1-\alpha_1 -\alpha_2 -\alpha_3)$ and ${\tilde G}^{\alpha \beta}={1 \over 2}\epsilon^{\alpha \beta \theta \tau}G_{\theta \tau}$.
The expressions of the LCDAs listed above, together with the numerical values of the parameters entering in such expressions, can be found in \cite{Ball:2006wn}. For the sake of clarity, we report below the correspondence between the LCDAs used in this paper and those in \cite{Ball:2006wn}:
\bea
&& \phi_K \to \phi_{2;K}\,\, ; \hskip 1 cm \phi_{4K} \to \phi_{4;K}\,\, ; \hskip 1 cm \psi_{4K} \to \psi_{4;K}\,\, ; \nonumber \\
&& \phi_P \to \phi^p_{3;K}\,\, ; \hskip 1 cm \phi_\sigma \to \phi^\sigma_{3,K}\,\, ; \hskip 1 cm \phi_{3K} \to \Phi^{3;K}\,\, ; \hskip 1 cm\nonumber \\
&&
\varphi_\parallel \to \Phi_{4;K}\,\, ; \hskip 1 cm \varphi_\perp \to \Psi_{4;K}\,\, ; \hskip 1 cm
{\tilde \varphi}_\parallel \to {\tilde \Phi}_{4;K}\,\, ; \hskip 1 cm {\tilde \varphi}_\perp \to {\tilde \Psi}_{4;K}\,\, .
\nonumber \eea

\end{document}